\documentstyle[psfig]{mn}

\newif\ifAMStwofonts

\ifoldfss
  \ifCUPmtlplainloaded \else
    \NewTextAlphabet{textbfit} {cmbxti10} {}
    \NewTextAlphabet{textbfss} {cmssbx10} {}
    \NewMathAlphabet{mathbfit} {cmbxti10} {} 
    \NewMathAlphabet{mathbfss} {cmssbx10} {} 
  \fi
  \ifAMStwofonts
    \ifCUPmtlplainloaded \else
      \NewSymbolFont{upmath} {eurm10}
      \NewSymbolFont{AMSa} {msam10}
      \NewMathSymbol{\upi}     {0}{upmath}{19}
      \NewMathSymbol{\umu}     {0}{upmath}{16}
      \NewMathSymbol{\upartial}{0}{upmath}{40}
      \NewMathSymbol{\leqslant}{3}{AMSa}{36}
      \NewMathSymbol{\geqslant}{3}{AMSa}{3E}

      \let\leq=\leqslant \let\leq=\leqslant
       
    \fi
  \fi
\fi 

\ifnfssone
  \newmathalphabet{\mathit}
  \addtoversion{normal}{\mathit}{cmr}{m}{it}
  \addtoversion{bold}{\mathit}{cmr}{bx}{it}
  \newmathalphabet{\mathbfit} 
  \addtoversion{normal}{\mathbfit}{cmr}{bx}{it}
  \addtoversion{bold}{\mathbfit}{cmr}{bx}{it}
  \newmathalphabet{\mathbfss} 
  \addtoversion{normal}{\mathbfss}{cmss}{bx}{n}
  \addtoversion{bold}{\mathbfss}{cmss}{bx}{n}
  \ifAMStwofonts
    \ifCUPmtlplainloaded \else
      \UseAMStwoboldmath
      \makeatletter
      \new@mathgroup\upmath@group
      \define@mathgroup\mv@normal\upmath@group{eur}{m}{n}
      \define@mathgroup\mv@bold\upmath@group{eur}{b}{n}
      \edef\UPM{\hexnumber\upmath@group}
      \new@mathgroup\amsa@group
      \define@mathgroup\mv@normal\amsa@group{msa}{m}{n}
      \define@mathgroup\mv@bold\amsa@group{msa}{m}{n}
      \edef\AMSa{\hexnumber\amsa@group}
      \makeatother
      \mathchardef\upi="0\UPM19
      \mathchardef\umu="0\UPM16
      \mathchardef\upartial="0\UPM40
      \mathchardef\leqslant="3\AMSa36
      \mathchardef\geqslant="3\AMSa3E

      \let\leq=\leqslant \let\leq=\leqslant

    \fi
  \fi
\fi 

\ifnfsstwo
  \DeclareMathAlphabet{\mathbfit}{OT1}{cmr}{bx}{it}
  \SetMathAlphabet\mathbfit{bold}{OT1}{cmr}{bx}{it}
  \DeclareMathAlphabet{\mathbfss}{OT1}{cmss}{bx}{n}
  \SetMathAlphabet\mathbfss{bold}{OT1}{cmss}{bx}{n}
  \ifAMStwofonts
    \ifCUPmtlplainloaded \else
      \DeclareSymbolFont{UPM}{U}{eur}{m}{n}
      \SetSymbolFont{UPM}{bold}{U}{eur}{b}{n}
      \DeclareSymbolFont{AMSa}{U}{msa}{m}{n}
      \DeclareMathSymbol{\upi}{0}{UPM}{"19}
      \DeclareMathSymbol{\umu}{0}{UPM}{"16}
      \DeclareMathSymbol{\upartial}{0}{UPM}{"40}
      \DeclareMathSymbol{\leqslant}{3}{AMSa}{"36}
      \DeclareMathSymbol{\geqslant}{3}{AMSa}{"3E}

      \let\leq=\leqslant \let\leq=\leqslant

    \fi
  \fi
\fi 

\ifCUPmtlplainloaded \else
  \ifAMStwofonts \else 
    \def\upi{\pi}
    \def\umu{\mu}
    \def\upartial{\partial}
  \fi
\fi

\title{Line temperatures and elemental abundances in HII galaxies}
\author[E. P{\'e}rez-Montero and A.I. D\'\i az]
       {Enrique~P{\'e}rez-Montero and Angeles I.~D{\'\i}az \\ 
        Departamento de F{\'\i}sica Te{\'o}rica, C-XI, Universidad Aut{\'o}noma de Madrid, 28049 Madrid, Spain\\}
\date{Accepted 
      Received ;
      in original form }

\pagerange{\pageref{firstpage}--\pageref{lastpage}}
\pubyear{2001}

\begin{document}

\maketitle

\label{firstpage}

\begin{abstract}

  We present long-slit spectrophotometric observations in the red and near infrared 
of 12 HII galaxies. The spectral range includes the sulphur lines [SII]
at $\lambda\lambda$ 6716, 6731 {\AA} and [SIII] at $\lambda$ 6312 {\AA} and $\lambda\lambda$ 9069, 9532
{\AA}. For all of the observed galaxies, at least three 
ion-weighted temperatures from forbidden auroral to nebular line
ratios have been obtained and the relations between
the different line temperatures have been discussed. It is found that,
for some objects, the [OII] temperatures derived from those of [OIII]
through the use of photo-ionisation models, without taking into account
the effect of density, can lead to a significant
underestimate of the O$^+$/H$^+$ ionic abundance and hence of the
total oxygen abundance.

For all the observed objects, we have calculated the ionic abundances of $O^+$, $O^{2+}$,
$S^+$, $S^{2+}$ and  $N^+$ and they have been used to constraint
the ionisation structure of the emmitting  regions with the help of
photo-ionisation models. From them, the ionisation correction
factors for N and S, and their corresponding total abundances,  have been derived.

\end{abstract}


\section{Introduction}

HII galaxies form part of a wider class of galaxies called blue compact
galaxies (BCG's). Their light is dominated by a few parsecs diameter
core in which an  episode of violent star formation (VSF) is taking
place. Their  emission line spectrum
is nearly identical to those from giant extragalactic HII regions
(GEHR's) and therefore they are
easy to detect in objective prisms surveys.

They are systems generally metal-deficient with respect to the sun,
ionised by OB star associations, and for a long time they have been
considered as very young objects
forming their first generation of stars and hence an important key for
the understanding of 
the mechanism of formation and evolution of primeval galaxies. It has
recently been found, however, that HII galaxies may show besides this
dominant young component, stellar
populations of intermediate-to-old ages (Schulte-Ladbeck et al. 1998).
  Their morphology, despite to be compact objects, is also varied and it is
not always spherical (Telles et al., 1997). 

The metallicity of HII galaxies is a parameter of recognized
importance when trying to characterize their evolutionary status 
and link them to other objects showing overlapping properties, like
Dwarf Irregular (dI) or Low Surface Brightness Galaxies (LSBG). It 
affects stellar evolution in different ways and its
knowledge is imperative for the interpretation of integrated colours
and ionised gas properties. Metal content is also at the base of
global relations like those existing or searched for with luminosity
(Hunter \& Hoffman 1999), gas mass fraction (Pagel 1997) and/or emission
line width (Melnick et al. 1987).

The spectra of HII galaxies, like the spectra of HII regions, are 
dominated by the properties of the continuum spectral energy
distribution of their ionising 
clusters and the physical conditions of the ionised gas. Therefore,
in principle, we can 
use the same methods developped for the study of HII regions to
determine the chemical composition of 
HII galaxies. Abundances of He, N, O, Ne, S, Ar etc can be found from
optical spectra and, with the exception of He, are
derived from collisionally excited emission lines whose intensity
depends exponentially on electron temperature. This temperature is
found from the ratio of auroral to nebular line intensities of the same
ion, usually [OIII]$\lambda$4363 {\AA}/($\lambda$4959 {\AA} +$\lambda$5007 {\AA}).
 
At present there are good quality data for more than 100 objects. 
They show abundances in the range 7.1 $\leq$
12+log(O/H) $\leq$ 8.3 . The oxygen abundance peaks slightly above 1/10 of
the solar value.  Selection effects may be responsible for this since 
emission line ground-based surveys select galaxies with intense
forbidden oxygen lines which are strongest for an oxygen abundance of about 1/10
solar. On the other hand, for the sake of accuracy, only
galaxies with measurable [OIII] $\lambda$ 4363 {\AA} line are included in most
analysis. This introduces a sharp cut to abundances at about 
12+log(O/H) $\simeq$ 8.5 as well as an important bias.
Auroral lines are 
intrisically weak and therefore in many cases (objects of moderate
excitation and/or low surface brightness) remain undetected. In
average, HII galaxies without measurements of the [OIII] $\lambda$ 4363 {\AA}
line show higher oxygen abundances and lower ionisation parameters than
HII galaxies for which measurements of this line exist (D{\'\i}az
1999). They also show average H$\beta$ luminosities lower by about a factor
of two, although the total flux range is comparable in the two
subsamples and they cluster around higher redshifts. 

Alternative abundance determinations consist of empirical
calibrations of strong emission lines which are easily observable. The 
$R_{23}$ parameter, defined as the sum of the [OII]$\lambda\lambda$
3727,29 {\AA} and the [OIII]]$\lambda\lambda$ 4959,5007 {\AA} lines (Pagel et al. 1979)
calibrated against oxygen abundance is 
probably the most widely used. This calibration presents however
several difficulties (see e.g. D{\'\i}az 2000, Pilyugin 2000) among them the
dependence of $R_{23}$ on the degree of ionisation and the
two-valued nature of the calibration, this latter one being most
important when dealing with HII galaxies since about 40\% of them have 
values of $R_{23}$ in the turn-over ill-defined region of the
calibration where objects with the same value of $R_{23}$ have oxygen
abundances which differ by up to an order of magnitude.

Recently, D{\'\i}az \& P{\'e}rez-Montero (2000) have presented an alternative empirical
abundance calibration based on the intensities of the far red sulphur
lines [SII] at $\lambda\lambda$ 6717,6731 {\AA} and [SIII] at $\lambda\lambda$ 9069,9532 {\AA}. Due to
the lower sensitive to temperature of these lines, this calibration
remains single-valued up to solar metallicities and therefore can
provide more accurate empirically derived abundances. This calibration
should be improved however at the low metallicity end with the
inclusion of more HII
galaxy data which constituted only 10\% of the total
number of objects in the original calibration. 

On the other hand, recent work on moderate excitation GEHR in spiral
galaxies (D{\'\i}az et al. 2000; Castellanos et al. 2002) has shown that it
is possible to detect and measure [SIII] temperatures in regions where
the [OIII]$\lambda$4363 {\AA} remains undetected. This could allow accurate abundance 
determinations for a large number of objects provided their ionisation
structure is understood.    

We have therefore undertaken a project for the observation and
measure of the sulphur lines in a sample of HII galaxies with the
two-folded aim of probing their ionisation structure, by comparing the
directly derived different ionic temperatures with model predictions,
and improving the oxygen abundance calibration through the
determination of the $S_{23}$ parameter for these objects.

The characteristics of the sample objects, together with the
description of the observations and reductions of the data are
given in Section 2. The results are presented in Section 3 and
discussed in Section 4. Finally, Section 5 summarizes the most
important conclusions of this work.

\section {Observations and data reduction}

The sample of observed objects consists of 12 HII galaxies
selected on the basis of their high $H\beta$ fluxes
($-13.60 \leq logF(H\beta) \leq -14.30$, $F(H\beta)$ in $ erg\cdot cm^{-2}\cdot s^{-1}$) and
low metallicities ($0.06 Z_{\odot} \leq Z \leq 0.2 Z_{\odot}$) as 
determined from direct measurements of the of [OIII] auroral line
at $\lambda$ 4363 {\AA} . Most of them belong to the first and second Byurakan 
objective prism surveys. The characteristics of the sample objects, together
with their references are given in Table 1.


\begin{table*}
\begin{minipage}{170mm}
\footnotesize
\caption{Characteristics of the observed sample galaxies}
\begin{center}
\begin{tabular}{cccclcccc}
\hline
\hline
Object & Ref.\footnote{References: (1){\em Guseva, Izotov \& Thuan (2000)},
(2){\em Izotov \& Thuan 1998 (IT98), (3){\em Izotov et al. 1997 (ITL97)} }, 
(4){\em Terlevich et al. 1991 (T91)}, (5) {\em Izotov et al. 1994 (ITL94)}} & 
\multicolumn{2}{c} {Coordinates
  (2000)}\footnote{Units of right ascenssion are hours, minutes and
  seconds, and units of declination are degrees, arcminutes and
  arcseconds.} & Other names &redshift & $-\log F(H{\beta})$ &
$12+\log(O/H)$ & $M_V$ \\
 & & $\alpha$  & $\delta$ & & & \\
\hline
0553+034 & 1 & 05 55 42.8 & +03 23 30 & IIZw40   & 0.00236 $\pm$ 0.00004 &
13.04 & 8.15 &   -- \\
0635+756 & 2 & 06 42 15.5 & +75 37 33 & Mrk5     & 0.00241 $\pm$ 0.00004 &
13.39 & 8.20 & -15.1\\
0749+568 & 3 & 07 53 41.1 & +56 41 51 &          & 0.01803 $\pm$ 0.00008 & 
13.72 & 7.89 & -16.3\\
0926+606 & 3 & 09 30 09.4 & +60 28 06 &          & 0.01340 $\pm$ 0.00006 & 
13.34 & 7.95 & -16.2\\
0946+171 & 4 & 09 49 18.0 & +16 52 46 & Mrk709   & 0.05162 $\pm$ 0.00006 & 
13.76 & 7.68 &   -- \\
0946+558 & 5 & 09 49 30.4 & +55 34 49 & Mrk22    & 0.00494 $\pm$ 0.00006 & 
13.68 & 8.04 & -15.8\\
1030+583 & 3 & 10 34 10.1 & +58 03 49 & Mrk1434  & 0.00726 $\pm$ 0.00006 & 
13.44 & 7.83 & -15.9\\
1102+294 & 2 & 11 04 58.5 & +29 08 22 & Mrk 36   & 0.00178 $\pm$ 0.00005 & 
13.26 & 7.81 & -13.6\\
1124+792 & 3 & 11.27.59.9 & +78.59.39 & VIIZw403 & -0 00058 $\pm$ 0.00009 &
13.33 & 7.73 & -13.7 \\
1148-020 & 2 & 11 51 33.0 & -02 22 23 & UM461    & 0.00310 $\pm$ 0.00004 &
13.47 & 7.80 & -13.8\\
1150-021 & 2 & 11 52 37.3 & -02 28 10 & UM462    & 0.00318 $\pm$ 0.00005 &
13.03 & 8.00 & -15.9\\
1223+487 & 3 & 12 26 16.0 & +48 29 37 & Mrk209, IZw36&0.00061 $\pm$ 0.00006 &
12.84 & 7.81 & -12.6\\
\hline
\end{tabular}
\end{center}
\end{minipage}
\end{table*}


\normalsize

The observations were made with the Isaac Newton Telescope 
(INT) in 1999 February and 2000 January at the Roque de los Muchachos 
Observatory using
the Intermediate Dispersion Spectrograph (IDS), the 235-mm camera and a 
TEK - CCD detector. A R600R grating was used to cover two different
spectral ranges of about 1700 {\AA} each, centered at $\lambda$ 7050 {\AA} and 
$\lambda$ 8850 {\AA}. The spectral dispersion of  1.66 {\AA}/pixel combined with
a slit width of 2 arcsec yielded a spectral resolution of about 5 {\AA}
FWHM. The spatial resolution is about 0.9 arcsec/pixel and the seeing
oscilated  between $1^{\prime \prime}$ and $1^{\prime \prime} .5$. A journal of
observations is given in Table 2.

Several  exposures were taken for each object thus allowing an adequate
removal of cosmic ray hits.


The data were reduced using the IRAF (Image Reduction and Analysis
Facility) package following standard procedures. Firstly, the two
dimensional spectra were bias subtracted and flat-field corrected. In order 
to do this we have used combined flat-fields taken before and after of each exposure. Wavelength calibration was achieved by means of comparison spectra of 
Ne-Ar lamps taken before and after the observation of each object.
In all cases the calibration was accurate to 0.1 {\AA}. 

One dimensional sky subtracted spectra were extracted using the
routine APALL. They were subsequently corrected for atmospheric extinction
using an appropriate extinction curve for La Palma Observatory. 
Finally, flux calibration was performed using observations of the
standard stars: HD19445 and HD93521.

The spectral regions around the
red [SIII] lines are heavily cut up by atmospheric water vapour
absorption bands. In our
observations this absorption is more important at wavelengths longer
that 9200 \AA. This bands were removed from our spectra by dividing by
the spectrum of a standard star normalized to unity and with the stellar
absorption lines supressed (see D\'\i az, Pagel \& Wilson 1985) . 
This method proved to be good in many cases
but in some of them  it was not entirely successful  and therefore the
[SIII] $\lambda$ 9532 \AA{} line measurements could not be
used with the necessary confidence. Fortunately, in all cases the
[SIII] $\lambda$ 9069 \AA{} line is relatively unaffected  by absorption.


\begin{table*}
\begin{minipage}{170mm}
\footnotesize
\caption{Journal of observations}
\begin{center}
\begin{tabular}{cccc}
\hline
\hline
Object & Night & $\lambda$ range  ({\AA}) & Exposure (s) \\
\hline
0553+034 & February 6/7 1999& 6200 - 7900 & 1$\times$1200 \\
0553+034 & February 6/7 1999& 8000 - 9700 & 2$\times$1200 \\
0635+756 & February 6/7 1999& 6200 - 7900 & 2$\times$1800 \\
0635+756 & February 6/7 1999& 8000 - 9700 & 2$\times$1800 \\
0749+568 & January 11/12 2000& 6200 - 7900 & 1$\times$1200 + 1$\times$1800\\
0749+568 & January 11/12 2000& 8000 - 9700 & 2$\times$1800 \\
0926+606 & February 6/7 1999& 6200 - 7900 & 2$\times$1800 \\
0926+606 & February 6/7 1999& 8000 - 9700 & 2$\times$1800 \\
0946+171 & January 11/12 2000& 6200 - 7900 & 2$\times$1800 \\
0946+171 & January 11/12 2000& 8000 - 9700 & 2$\times$1800 \\
0946+558 & February 7/8 1999& 6200 - 7900 & 2$\times$1800 \\
0946+558 & February 7/8 1999& 8000 - 9700 & 2$\times$1800 \\
1030+583 & February 7/8 1999& 6200 - 7900 & 2$\times$1800 \\
1030+583 & February 7/8 1999& 8000 - 9700 & 2$\times$1800 \\
1102+294 & January 11/12 2000& 6200 - 7900 & 2$\times$1200 \\
1102+294 & January 11/12 2000& 8000 - 9700 & 1$\times$1462 \\
1124+792 & February 7/8 1999& 6200 - 7900 & 2$\times$1800 \\
1124+792 & February 7/8 1999& 8000 - 9700 & 2$\times$1800 \\
1148-020 & February 8/9 1999& 6200 - 7900 & 1$\times$1800 + 1$\times$500 \\
1148-020 & February 8/9 1999& 8000 - 9700 & 2$\times$1800 \\
1150-021 & February 8/9 1999& 6200 - 7900 & 1$\times$1800 + 1$\times$1200 \\
1150-021 & February 8/9 1999& 8000 - 9700 & 3$\times$1200 \\
1223+487 & February 6/7 1999& 6200 - 7900 & 2$\times$1800 \\
1223+487 & February 6/7 1999& 8000 - 9700 & 2$\times$1800 \\

\hline
\end{tabular}
\end{center}

\end{minipage}
\end{table*}


\section{Results}

Representative spectra of two of the observed galaxies: IIZw40 and 
UM461 in the red and near IR spectral ranges are shown in Figures
1 and 2 with the identification of the  lines relevant to the determination of the sulphur abundances.


\begin{figure*}
\setcounter{figure}{0}
\begin{minipage}{150mm}
\psfig{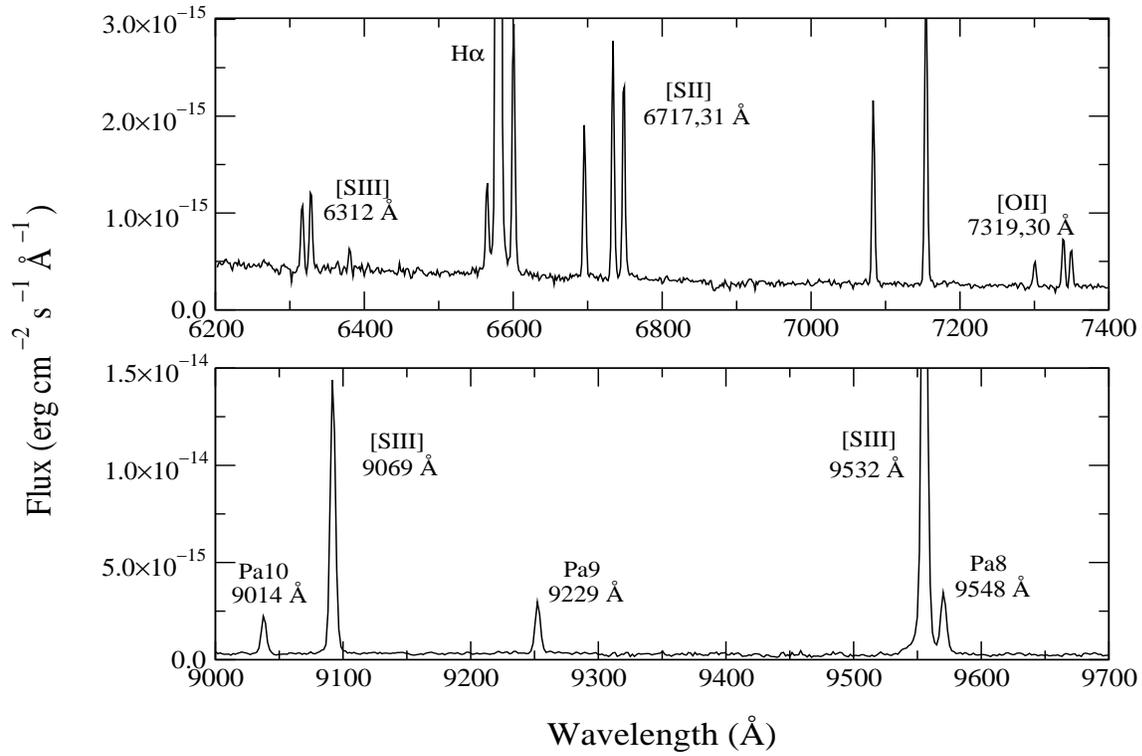}
\caption{Spectra of IIZw40 in the two observed ranges}
\end{minipage}
\end{figure*} 


\begin{figure*}
\setcounter{figure}{1}
\begin{minipage}{150mm}
\psfig{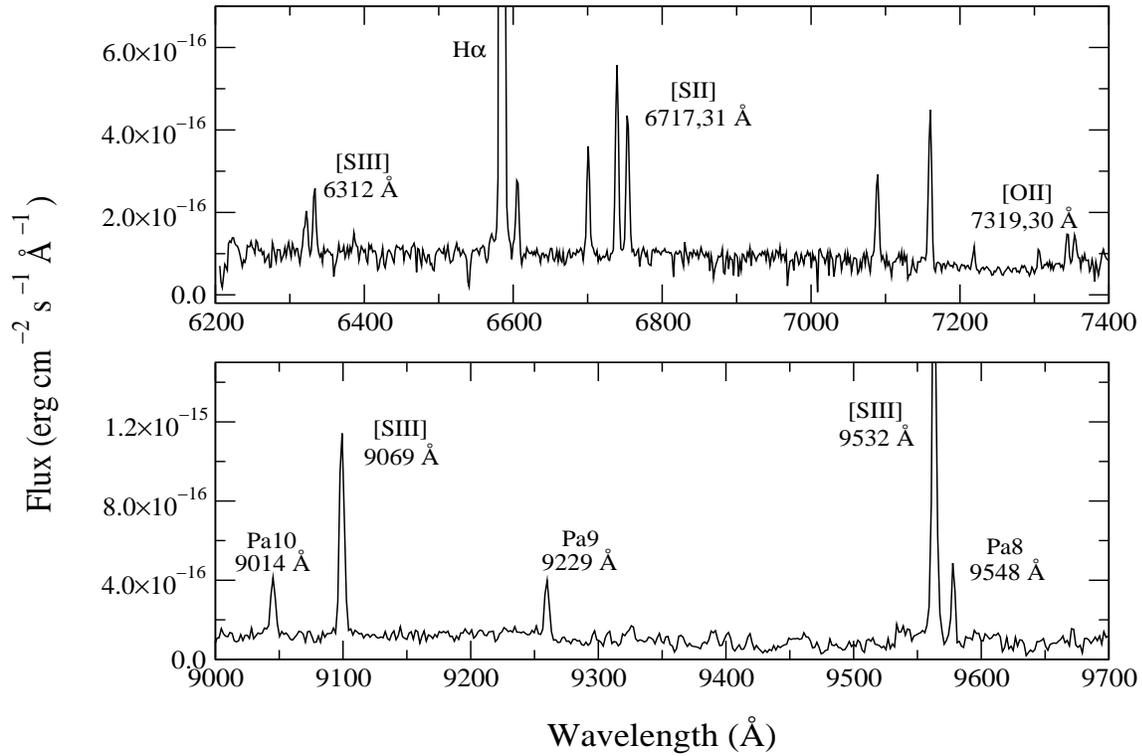}
\caption{Spectra of UM461 in the two observed ranges}
\end{minipage}
\end{figure*} 


\subsection{Emission lines intensities}

The measurement of the emission line fluxes was made using the
SPLOT routine, that integrates the line intensity over a local-fitted
continuum. The errors in
the line fluxes have been calculated from the expression
$\sigma_{l}$ = $\sigma_{c}$N$^{1/2}$[1 + EW/(N$\Delta$)]$^{1/2}$, where
$\sigma_{l}$ is the error in the line flux, $\sigma_{c}$ represents the
standard deviation in a box near the measured emission line and stands for the
error in the continuum placement, N is the number of pixels used in the
measurement of the line flux, EW is the line equivalent width, and
$\Delta$ is the wavelength dispersion in angstroms per pixel.

We have not attempted to quantify the amount of internal
reddening, but have rather decided to measure each line intensity
relative to that of the nearest hydrogen recombination line which, in
turn, has been
taken to be equal to its theoretical Case B recombination value (e.g. 
Osterbrock 1989) at the electron temperature estimated from the observed [OIII]
lines ratio. This procedure also minimizes any possible error 
introduced by the  
flux-calibration. Only the [OII] $\lambda\lambda$ 7319, 7330 {\AA} line intensities have been corrected for
reddening for which published values of the reddening constant C(H$\beta$)
have been used. In the  cases in which the [SIII] $\lambda$ 9532 \AA{} line could not be
measured with the required level of confidence due to poor water absorption
corrections, only the [SIII] $\lambda$ 9069 \AA{} line was used and a ratio
between the two lines of 2.44 was assumed.

Our measured emission line intensities relative to H$\beta$ = 100, are given in Table 3 
together with data on the blue and visible spectral ranges taken from
the literature. For most of the objects, the comparison of the intensities
of the emission lines in the  
overlapping spectral region shows an acceptable agreement.


\begin{table*}
\begin{minipage}{150mm}
\small
\caption{{\large Relative emission line intensities}}
\begin{center}
\begin{tabular}{ccccccc}
\hline
\hline
$\lambda$  ({\AA}) & \multicolumn{2}{c}{0553+034 $\equiv$ II Zw 40} & \multicolumn{2}{c}{0635+756 $\equiv$ Mrk 5} & \multicolumn{2}{c}{0749+568}   \\
 & GIT00 & our data & IT98 & our data & ITL97 & our data \\
\hline
3727 [OII] & 83.9$\pm$1.2  &   & 212.9$\pm$3.1  &   & 166.8$\pm$4.3  &\\
4072 [SII] &   &   &   &   & 2.0$\pm$1.1&\\
4363 [OIII] & 10.9$\pm$0.4  &   & 4.4$\pm$0.5  &   & 9.8$\pm$1.1  &\\
4861 H$\beta$ & 100.0$\pm$1.0  & 100  & 100.0$\pm$1.5  & 100  & 100.0$\pm$2.7  &
100 \\
4959 [OIII] & 246.2$\pm$2.1  &   & 129.8$\pm$1.8  &   & 167.2$\pm$3.9  &\\
5007 [OIII] & 740.9$\pm$5.6  &   & 381.5$\pm$4.5  &   & 488.0$\pm$9.9  &\\
5199 [NI] &   &   &   &   &  &\\
6300 [OI] & 1.6$\pm$0.1 & 1.23$\pm$0.04 & 4.3$\pm$0.3 & 3.4$\pm$0.2 & 4.1$\pm$0.8 & 5.3$\pm$0.2 \\
6312 [SIII] & 1.6$\pm$0.1 & 1.39$\pm$0.04 & 2.1$\pm$0.4 & 1.6$\pm$0.2 & 1.8$\pm$0.6 & 2.3$\pm$0.3 \\
6363 [OI] &  & 0.42$\pm$0.04 & 1.3$\pm$0.3 & 1.8$\pm$0.2 & 2.3$\pm$0.1 & 1.9$\pm$0.2  \\
6548 [NII] &  & 2.82$\pm$0.04 &  &5.4$\pm$0.2 &  &2.4$\pm$0.3 \\
6563 H$\alpha$ & 287.2$\pm$2.4 & 281.0$\pm$0.9 & 285.7$\pm$3.7 & 283.0$\pm$2.6 & 279.7$\pm$6.4 &280.0$\pm$1.7\\
6584 [NII] & 6.3$\pm$0.2 & 5.8$\pm$0.1 & 13.8$\pm$0.5 & 13.7$\pm$0.2 & 7.6$\pm$0.7 & 6.8$\pm$0.1\\
6678 [HeI] & 3.2$\pm$0.1 & 3.2$\pm$0.1 & 3.1$\pm$0.3 & 3.5$\pm$0.1 & 3.9$\pm$0.8 & 2.7$\pm$0.1\\
6717 [SII] & 6.7$\pm$0.2 & 6.09$\pm$0.08 & 23.3$\pm$0.6 & 21.9$\pm$0.1 & 17.8$\pm$1.1 & 13.3$\pm$0.1\\
6731 [SII] & 5.4$\pm$0.1 & 5.17$\pm$0.08 & 16.6$\pm$0.5 & 15.7$\pm$0.1 & 11.4$\pm$0.8 & 10.1$\pm$0.2\\
7065 [HeI] & 4.2$\pm$0.1  &4.9$\pm$0.1 & 2.5$\pm$0.2 & 2.1$\pm$0.1 & 2.4$\pm$0.5 & 2.2$\pm$0.2\\
7137 [ArIII] & 7.7$\pm$0.1 &9.1$\pm$0.1 & 8.7$\pm$0.3 & 8.7$\pm$0.2 & 6.5$\pm$0.7 & 5.1$\pm$0.1\\
7319 [OII] &  &1.57$\pm$0.08 &  &2.9$\pm$0.2 & 3.1$\pm$0.6  &1.9$\pm$0.1 \\
7330 [OII] & &1.30$\pm$0.08 &  &2.7$\pm$0.2 & 1.8$\pm$0.4  &1.5$\pm$0.1\\
9014 Pa10 &  &1.8$\pm$0.1 &  &1.8$\pm$0.1 &  &1.8$\pm$0.2  \\
9069 [SIII] &  &13.9$\pm$0.6 &  &15.9$\pm$1.2 & &12.7$\pm$1.4  \\
9532 [SIII] &  &27.9$\pm$1.0 &  &32.2$\pm$2.2 &  &  \\
9548 Pa8 &  &3.5$\pm$0.1 &  &3.6$\pm$0.2 &  & \\
\hline
C(H$\beta$) & 0.09 & & 0.42 & & 0.12 & \\
\hline
\end{tabular}
\end{center}
\end{minipage}
\end{table*}


\begin{table*}
\begin{minipage}{150mm}
\small
\setcounter{table}{2}
\caption{{\large Continued}}
\begin{center}
\begin{tabular}{ccccccc}
\hline
\hline
$\lambda$  ({\AA})& \multicolumn{2}{c}{0926+606} &  \multicolumn{2}{c}{0946+171 $\equiv$ Mrk 709 } &\multicolumn{2}{c}{0946+558 $\equiv$ Mrk 22}  \\
 & ITL97 & our data & T91 & our data & ITL94& our data   \\
\hline
3727 [OII]     &178.5$\pm$1.2 &  & 183.6$\pm$18.1  & & 148.7$\pm$2.3 &                    \\
4072 [SII]     &2.6$\pm$0.4   &  &  &   & 1.5$\pm$0.3   &                \\
4363 [OIII]    &8.3$\pm$0.3   &  &  8.8$\pm$0.5   &  &  8.2$\pm$0.3   &                \\
4861 H$\beta$  &100.0$\pm$0.7 & 100  & 100  & 100  & 100.0$\pm$1.1 & 100                 \\
4959 [OIII]    &162.8$\pm$1.0 &  & 121.5$\pm$0.9  & &182.4$\pm$1.6 &                    \\
5007 [OIII]    &477.2$\pm$2.6 &  & 369.6$\pm$4.2 & &545.5$\pm$4.2 &                    \\
5199 [NI]      &0.7$\pm$0.2   &             &     &  & 0.3$\pm$0.1   &                 \\
6300 [OI]      &3.6$\pm$0.2   & 3.6$\pm$0.2 &             & 7.3$\pm$0.1 & 1.9$\pm$0.1 & 1.7$\pm$0.1 \\
6312 [SIII]    &1.9$\pm$0.2   & 2.0$\pm$0.1 &             & 1.4$\pm$0.1 & 2.4$\pm$0.1 & 3.0$\pm$0.1  \\
6363 [OI]      &1.1$\pm$0.2   & 1.4$\pm$0.1 &             & 2.5$\pm$0.2 & 0.6$\pm$0.1 &  1.1$\pm$0.3      \\
6548 [NII]     &              & 3.3$\pm$0.2 &               &   8.5$\pm$0.1    &     & 2.6$\pm$0.3    \\
6563 H$\alpha$ &280.4$\pm$1.7 & 282.0$\pm$0.7 & 286.0$\pm$ 1.0  & 281.0$\pm$0.7 & 282.5$\pm$2.5 & 283.0$\pm$1.3  \\
6584 [NII]     &8.3$\pm$0.4   & 9.3$\pm$0.4 & 27.9$\pm$2.2 & 34.2$\pm$0.2 & 6.4$\pm$0.2 & 8.0$\pm$0.1  \\
6678 [HeI]     &3.0$\pm$0.2   & 3.3$\pm$0.4 &             & 3.1$\pm$0.2 & 2.9$\pm$0.2 & 3.4$\pm$0.1  \\
6717 [SII]     &18.2$\pm$0.3  & 18.7$\pm$0.3 &31.3$\pm$3.0 &36.2$\pm$0.2 & 11.4$\pm$0.3 & 15.1$\pm$0.1\\
6731 [SII]     &14.6$\pm$0.3  & 13.3$\pm$0.2 &28.5$\pm$3.0 &26.1$\pm$0.2& 8.5$\pm$0.2  & 11.5$\pm$0.4  \\
7065 [HeI]     &2.2$\pm$0.1   & 2.2$\pm$0.1 &             & 2.1$\pm$0.1& 2.4$\pm$0.2 & 2.7$\pm$0.3  \\
7137 [ArIII]   &6.6$\pm$0.2   & 7.4$\pm$0.1 &             & 8.0$\pm$0.1& 6.5$\pm$0.2 & 6.8$\pm$0.2  \\
7319 [OII]     &2.6$\pm$0.2   & 2.2$\pm$0.1 &             & 2.9$\pm$0.1& 1.6$\pm$0.2 & 1.0$\pm$0.2  \\
7330 [OII]     &2.0$\pm$0.1   & 2.1$\pm$0.1 &             & 2.8$\pm$0.1& 2.0$\pm$0.2 & 1.6$\pm$0.1  \\
9014 Pa10      &              & 1.8$\pm$0.1 &               &   1.8$\pm$0.1  &     &1.8$\pm$0.3     \\
9069 [SIII]    &              & 12.2$\pm$1.2 &              &  8.7$\pm$1.0 &           &14.8$\pm$2.0    \\
9532 [SIII]    &              & 36.0$\pm$4.6&              &               &           &    \\
9548 Pa8       &              & 3.5$\pm$0.1 &              &              &            &     \\
\hline
C(H$\beta$)    & 0.29         &  & 0-0.22    &     & 0.18        &                \\
\hline
\end{tabular}
\end{center}
\end{minipage}
\end{table*}

\begin{table*}
\begin{minipage}{150mm}
\small
\setcounter{table}{2}
\caption{{\large Continued}}
\begin{center}
\begin{tabular}{ccccccc}
\hline
\hline
$\lambda$  ({\AA}) &  \multicolumn{2}{c}{1030+583 $\equiv$ Mrk 1434} &  \multicolumn{2}{c}{1102+294 $\equiv$ Mrk 36} & \multicolumn{2}{c}{1124+792 $\equiv$ VII Zw 403} \\
 & ITL97& our data & IT98 & our data  & ITL97 & our data \\
\hline
3727 [OII]     & 96.8$\pm$0.6   &  & 129.3$\pm$1.5  &  & 133.3$\pm$0.9  &  \\
4072 [SII]     & 1.8$\pm$0.3    &  & 1.6$\pm$0.5  &  &1.3$\pm$0.4  &   \\
4363 [OIII]    & 10.4$\pm$0.2   &    & 9.6$\pm$0.5 &  & 7.1$\pm$0.2 & \\
4861 H$\beta$  & 100.0$\pm$0.6  & 100  & 100.0$\pm$1.1  & 100  &100.0$\pm$0.7  & 100 \\
4959 [OIII]    & 170.4$\pm$0.8  &    & 162.2$\pm$1.6  &  &114.3$\pm$0.8  & \\
5007 [OIII]    & 502.8$\pm$2.1  &     & 483.4$\pm$4.2  &  &345.5$\pm$3.9  & \\
5199 [NI]      & 0.8$\pm$0.2    &     &     &    &\\
6300 [OI]      & 2.6$\pm$0.2    &   1.4$\pm$0.1  & 2.8$\pm$0.3 & 2.0$\pm$0.2& 1.9$\pm$0.1 &\\
6312 [SIII]    & 1.8$\pm$0.1    &   1.6$\pm$0.1& 1.8$\pm$0.3 & 1.8$\pm$0.2 & 1.3$\pm$0.1 & 1.3$\pm$0.1\\
6363 [OI]      & 0.6$\pm$0.1    &   0.7$\pm$0.1& 1.2$\pm$0.2 & 0.9$\pm$0.2 & 1.3$\pm$0.1 \\
6548 [NII]     &                &   1.4$\pm$0.2 &   & 1.7$\pm$0.1&  & 1.9$\pm$0.3\\
6563 H$\alpha$ & 278.6$\pm$1.3  & 281.0$\pm$0.3 & 279.1$\pm$2.7 & 281.0$\pm$0.9 & 278.7$\pm$1.6 & 281.0$\pm$0.9\\
6584 [NII]     &  3.1$\pm$0.1   &   3.9$\pm$0.3 & 5.3$\pm$0.3 & 4.9$\pm$1.2& 5.0$\pm$0.2 & 4.2$\pm$0.2\\
6678 [HeI]     & 2.8$\pm$0.1    &   2.7$\pm$0.1 & 2.7$\pm$0.3 &  2.6$\pm$0.1& 2.9$\pm$0.1 & 2.9$\pm$0.1\\
6717 [SII]     &  9.6$\pm$0.2   &   9.04$\pm$0.08 & 11.7$\pm$0.4 & 11.4$\pm$0.1& 10.3$\pm$0.2 & 7.7$\pm$0.1\\
6731 [SII]     &  6.7$\pm$0.1   &   6.03$\pm$0.08 &  8.4$\pm$0.4 &  8.6$\pm$0.1 &  7.5$\pm$0.2 & 5.3$\pm$0.1\\
7065 [HeI]     & 2.3$\pm$0.1    &   2.1$\pm$0.1 & 2.5$\pm$0.3 & 2.1$\pm$0.2& 2.0$\pm$0.1 &\\
7137 [ArIII]   & 6.0$\pm$0.1    &   4.8$\pm$0.1 & 6.2$\pm$0.3 & 5.4$\pm$0.2 & 6.4$\pm$0.2 & 5.5$\pm$0.1\\
7319 [OII]     & 1.8$\pm$0.2    &   1.3$\pm$0.1 &  & 2.3$\pm$0.2 & 2.3$\pm$0.2  & 1.3$\pm$0.1\\
7330 [OII]     & 1.0$\pm$0.1    &   1.0$\pm$0.1  &  & 1.6$\pm$0.2& 1.5$\pm$0.2  & 0.9$\pm$0.1\\
9014 Pa10      &                &   1.8$\pm$0.3  &  &1.8$\pm$0.1&  &1.8$\pm$0.2 \\
9069 [SIII]    &                &   9.2$\pm$0.8 & & 11.9$\pm$1.4 & &11.7$\pm$1.3\\
9532 [SIII]    &                &        &  &29.5$\pm$3.5 &  & \\
9548 Pa8       &                &    &  & 3.5$\pm$0.4&  & \\
\hline
C(H$\beta$)      & 0.00         &   & 0.02 &   & 0.00 &    \\
\hline
\end{tabular}
\end{center}
\end{minipage}
\end{table*}
%
\begin{table*}
\begin{minipage}{150mm}
\small
\setcounter{table}{2}
\caption{{\large Continued}}
\begin{center}
\begin{tabular}{ccccccc} 
\hline
\hline
$\lambda$  ({\AA})& \multicolumn{2}{c}{1148-020  $\equiv$ UM 461} & \multicolumn{2}{c}{1150-021 $\equiv$ UM 462} & \multicolumn{2}{c}{1223+487 $\equiv$ Mrk 209}  \\
  & IT98 & our data & IT98 & our data & ITL97& our data\\
\hline
3727 [OII] &   52.7$\pm$1.5  &   & 174.2$\pm$1.0&   &71.9$\pm$0.2  &  \\

4072 [SII] &   & & 2.2$\pm$0.4 &  &1.2$\pm$0.1  &   \\

4363 [OIII] &  13.6$\pm$0.7  &   & 7.8$\pm$0.2 &  & 12.7$\pm$0.1  &  \\

4861 H$\beta$ &  100.0$\pm$2.0  & 100  & 100.0$\pm$0.6& 100  &100.0$\pm$0.2 & 100    \\

4959 [OIII] & 203.9$\pm$3.4  &   & 166.3$\pm$0.9&  &196.0$\pm$0.3  &    \\
5007 [OIII] &  602.2$\pm$9.0  &   & 492.9$\pm$2.3&  &554.3$\pm$0.8  &   \\
5199 [NI] &   &   &  &   &0.3$\pm$0.1  &    \\

6300 [OI] &  1.9$\pm$0.4 & 1.7$\pm$0.1 & 3.8$\pm$0.2 & 3.5$\pm$0.2 & 1.4$\pm$0.1 & 1.6$\pm$0.2 \\
6312 [SIII] & 1.2$\pm$0.4 & 2.0$\pm$0.1 & 1.9$\pm$0.1 & 2.1$\pm$0.2 & 1.7$\pm$0.1 & 1.9$\pm$0.1  \\
6363 [OI] &  &  & 1.2$\pm$0.1  & 1.1$\pm$0.1 & 0.5$\pm$0.1 &   \\
6548 [NII] &     &   &  & 3.5$\pm$0.3 &          & 1.6$\pm$0.2  \\
6563 H$\alpha$ &  278.4$\pm$4.8 & 280.0$\pm$0.7 & 282.6$\pm$1.5 & 282.0$\pm$0.6 & 277.7$\pm$0.5 & 280.0$\pm$0.1 \\
6584 [NII] &  2.1$\pm$0.4 &  2.8$\pm$0.2 & 7.3$\pm$0.2 & 7.3$\pm$0.2 & 2.9$\pm$0.1 & 3.6$\pm$0.1  \\
6678 [HeI] &  2.9$\pm$0.3 & 3.2$\pm$0.1 & 3.0$\pm$0.1 &  3.0$\pm$0.1 & 2.9$\pm$0.1 & 3.3$\pm$0.2 \\
6717 [SII] &  5.2$\pm$0.4 &  5.8$\pm$0.1 & 16.8$\pm$0.2 & 16.2$\pm$0.3 & 6.1$\pm$0.1 &  5.9$\pm$0.2  \\
6731 [SII]  &  4.2$\pm$0.4 &  4.5$\pm$0.1 & 11.2$\pm$0.2 & 12.1$\pm$0.2& 4.5$\pm$0.1 &  4.7$\pm$0.2\\
7065 [HeI] & 2.8$\pm$0.3 & 2.7$\pm$0.1 & 3.0$\pm$0.1 & 2.0$\pm$0.1 & 2.4$\pm$0.1 & 2.2$\pm$0.1 \\
7137 [ArIII]  & 4.6$\pm$0.4 & 4.8$\pm$0.1 & 8.5$\pm$0.1 & 5.7$\pm$0.1 & 5.9$\pm$0.1  &5.5$\pm$0.1\\
7319 [OII]  &  &1.0$\pm$0.1 &  & 2.34$\pm$0.06  & 1.1$\pm$0.1  &0.9$\pm$0.1\\
7330 [OII] &  &1.1$\pm$0.1 &  & 1.96$\pm$0.09 &  1.0$\pm$0.1 &0.8$\pm$0.1 \\
9014 Pa10 &  &1.8$\pm$0.2 &  &1.8$\pm$0.2 &  &1.8$\pm$0.1  \\
9069 [SIII]  &  & 12.4$\pm$0.8 & & 10.5$\pm$1.5 &  &12.2$\pm$1.2 \\
9532 [SIII]  &  &21.7$\pm$ 0.7 &  &35.0$\pm$3.6 &  & \\
9548 Pa8 &  &3.5$\pm$0.3 &  &3.5$\pm$0.4 &  &  \\
\hline
C(H$\beta$)  & 0.12 & & 0.29 & & 0.00 &\\
\hline
\end{tabular}
\end{center}
\end{minipage}
\end{table*}


\subsection{Line temperatures, number densities and ionic abundances}

Electron densities were determined from the [SII] $\lambda$6717 {\AA}/$\lambda$6731 {\AA}
ratio using a five-level statistical equilibrium model (De Robertis,
Dufour \& Hunt 1987; Shaw \& Dufour 1995). Except in IIZw40, with a value
290 $cm^{-3}$, the diagnostic ratios give low values of the
density. 

For all the objects we have derived the value of t([SIII]),
from the
($\lambda$9069{\AA}+$\lambda$9532{\AA})/$\lambda$6312{\AA} ratio. Besides, we have determined t([OII]), from the ratio of
$\lambda$3727{\AA}/($\lambda$7319{\AA}+$\lambda$7330{\AA}), by combining, in most cases, our data in the red
with complementary data in the blue taken from the literature (see
Table 3). The same sources of data have been used to derive t([OIII]), from the ratio of
($\lambda$4959{\AA}+$\lambda$5007{\AA})/$\lambda$4363{\AA} and, for 7 objects of the sample, t([SII])
from the ratio $\lambda$(6717{\AA}+$\lambda$6731{\AA})/$\lambda$4072{\AA}. In the cases in which the agreement between
our measured line intensities and those in the literature is shown to
be good,  the values with the smaller observational
errors have been used to derive a
given line temperature. Otherwise we have preferred to use the line intensities
corresponding to the same data source.   In all cases, line
temperatures were derived from the corresponding line ratios using the
five-level atom code, assuming in each case the electron density
deduced from the [SII] line ratio and using the most recent  atomic collision
strength data as listed in Table 4.

The [OII] $\lambda$7319{\AA}+$\lambda$7330{\AA} lines as well as  the [SII]$\lambda$$\lambda$ 4068,
4074 \AA{} lines can have a contribution by direct recombination which
increases with temperature. We have estimated these contributions using
the calculated [OIII] electron temperatures to be less than 4 \% for
the [OII] lines and less than 0.01 \% for the [SII] lines in all cases
and therefore we have not corrected for this effect.


\begin{table}
\begin{minipage}{83mm}
\footnotesize
\setcounter{table}{3}
\caption{Sources of the effective collision strengths of each ion}
\begin{center}
\begin{tabular}{cc}
\hline
\hline
Ion & References \\
\hline
$[OII]$ & McLaughlin \& Bell, 1998 \footnote{and private communication} \\ 
$[OIII]$,$[NII]$ & Lennon \& Burke, 1994\\
$[SII]$ & Ramsbotton, Bell \& Stafford, 1996\\
$[SIII]$ & Tayal \& Gupta, 1999\\
\hline
\end{tabular}
\end{center}
\end{minipage}
\end{table}


For all the observed galaxies, at least three line temperatures have
been derived. Electron densities and line temperatures, with their
corresponding errors, for each of the
observed galaxies are given in Table 5.



\begin{table*}
\begin{minipage}{150mm}
\footnotesize
\setcounter{table}{4}
\caption{{\large Physical properties of the observed galaxies}}
\begin{center}
\begin{tabular}{lcccccc}
\hline
\hline
Object & II Zw 40 & Mrk 5 & 0749+568 & 0926+606 & Mrk 709 & Mrk 22\\
\hline

$\log U$ & -2.23$\pm$0.08 & -2.6$\pm$0.2 & -2.55$\pm$0.10 & -2.6$\pm$0.1 & -2.6$\pm$0.1 & -2.45$\pm$0.10 \\
$n_e$([SII]) & 290$\pm$60  & 20$\pm$10 & 100$\pm$50 & $\leq$50 & $\leq$30 & 110$\pm$80\\
$t_e$([OII]) & 1.27$\pm$0.06  &1.32$\pm$0.08  & 1.38$\pm$0.12 & 1.23$\pm$0.04 & 1.50$\pm$0.16 & 1.16$\pm$0.09\\
$t_e$([OIII])& 1.34$\pm$0.03 & 1.22$\pm$0.06 & 1.54$\pm$0.10 &1.43$\pm$0.03 & 1.67$\pm$0.06 & 1.35$\pm$0.03\\
$t_e$([SII]) & --  & -- & -- & 1.04$\pm$0.14 &  -- & 0.96$\pm$0.16\\
$t_e$([SIII])& 1.30$\pm$0.05 &1.33$\pm$0.15  & 1.86$\pm$0.36 & 1.52$\pm$0.18&  1.65$\pm$0.23 & 1.98$\pm$0.28\\
 & & & & \\
$O^+/H^+$ $(\times 10^4)$& 0.12$\pm$0.02  & 0.28$\pm$0.07 & 0.19$\pm$0.17 & 0.29$\pm$0.04  & 0.16$\pm$0.08 & 0.31$\pm$0.11 \\
$O^{2+}/H^+$ $(\times 10^4)$& 1.08$\pm$0.08  & 0.73$\pm$0.12  & 0.51$\pm$0.10 & 0.59$\pm$0.04  & 0.31$\pm$0.03 & 0.78$\pm$0.05\\
$12+\log(O/H)$& 8.08$\pm$0.03  & 8.00$\pm$0.08  & 7.84$\pm$0.09 & 7.95$\pm$0.04 & 7.68$\pm$0.09 & 8.04$\pm$0.06 \\
 & & & & \\
$S^+/H^+$ $(\times 10^6)$& 0.15$\pm$0.02  & 0.48$\pm$0.06  & 0.27$\pm$0.05 & 0.68$\pm$0.27  & 0.62$\pm$0.13 & 0.71$\pm$0.40  \\
$S^{2+}/H^+$ $(\times 10^6)$& 0.94$\pm$0.09   & 1.06$\pm$0.27 & 0.66$\pm$0.25 & 0.89$\pm$0.27  & 0.50$\pm$0.16 & 0.71$\pm$0.23  \\
&&&&\\
$\log(N^+/O^+)$ & -1.31$\pm$0.05 & -1.29$\pm$0.05 & -1.48$\pm$0.08 & -1.44$\pm$0.04 & -0.82$\pm$0.09 & -1.47$\pm$0.07\\
$N^+/H^+$ $(\times 10^6)$  & 0.59$\pm$0.06 & 1.24$\pm$0.15 & 0.56$\pm$0.10 & 0.91$\pm$0.09 & 2.33$\pm$0.39 & 0.84$\pm$0.15\\
$12+\log(N/H)$ & 6.77$\pm$0.07 & 6.71$\pm$0.12 & 6.36$\pm$0.17 & 6.51$\pm$0.08 & 6.86$\pm$0.08 & 6.57$\pm$0.13\\
\hline
\end{tabular}
\end{center}
\end{minipage}
\end{table*}


\begin{table*}
\begin{minipage}{150mm}
\footnotesize
\setcounter{table}{4}
\caption{{\large Continued}}
\begin{center}
\begin{tabular}{lcccccc}
\hline
\hline
Object & Mrk 1434 & Mrk 36 & VII Zw 403  & UM 461 & UM 462 & Mrk 209\\
\hline
$\log U$ & -2.4$\pm$0.1 & -2.65$\pm$0.20 & -2.6$\pm$0.1  & -2.1$\pm$0.1 & -2.6$\pm$0.1 & -2.3$\pm$0.1\\
$n_e$([SII]) & $\leq$10 &  90$\pm$40 & $\leq$10 & 140$\pm$70 &  80$\pm$60 & 190$\pm$140\\
$t_e$([OII]) & 1.24$\pm$0.08 & 1.37$\pm$0.12 & 1.42$\pm$0.12  & 1.64$\pm$0.18 & 1.19$\pm$0.03 & 1.22$\pm$0.08 \\
$t_e$([OIII])& 1.55$\pm$0.02 & 1.53$\pm$0.05 & 1.52$\pm$0.03  & 1.62$\pm$0.05 & 1.38$\pm$0.02 & 1.62$\pm$0.01\\
$t_e$([SII]) & 1.54$\pm$0.32 & 1.09$\pm$0.28 & 1.05$\pm$0.27   & --  & 1.00$\pm$0.15 & 1.25$\pm$0.13\\
$t_e$([SIII])& 1.73$\pm$0.20 & 1.62$\pm$0.30 & 1.30$\pm$0.14   & 1.95$\pm$0.16 & 1.66$\pm$0.25 & 1.60$\pm$0.17\\
 & & & & \\
$O^+/H^+$ $(\times 10^4)$& 0.16$\pm$0.04  & 0.15$\pm$0.05 & 0.14$\pm$0.04  & 0.04$\pm$0.01  & 0.32$\pm$0.03 & 0.12$\pm$0.03\\
$O^{2+}/H^+$ $(\times 10^4)$& 0.51$\pm$0.02  & 0.50$\pm$0.05 & 0.36$\pm$0.02  & 0.55$\pm$0.05  & 0.66$\pm$0.03 & 0.51$\pm$0.01 \\
12+log(O/H)& 7.82$\pm$0.04  & 7.82$\pm$0.06 & 7.70$\pm$0.05  & 7.77$\pm$0.05  & 8.00$\pm$0.03 & 7.80$\pm$0.03 \\
& & & & \\
$S^+/H^+$ $(\times 10^6)$& 0.15$\pm$0.07  & 0.43$\pm$0.37 & 0.30$\pm$0.28  & 0.09$\pm$0.02  & 0.67$\pm$0.32 & 0.15$\pm$0.04  \\
$S^{2+}/H^+$ $(\times 10^6)$& 0.51$\pm$0.12   & 0.72$\pm$0.29 & 0.92$\pm$0.28  & 0.47$\pm$0.06  & 0.76$\pm$0.25 & 0.73$\pm$0.18 \\
&&&&\\
$\log(N^+/O^+)$ & -1.55$\pm$0.08 & -1.52$\pm$0.07 & -1.55$\pm$0.08 & -1.27$\pm$0.09 & -1.51$\pm$0.09 & -1.48$\pm$0.06\\
$N/H^+$ $(\times 10^6)$  & 0.38$\pm$0.08 & 0.38$\pm$0.15 & 0.36$\pm$0.08 & 0.18$\pm$0.05 & .0.82$\pm$0.17 & 0.38$\pm$0.07\\
$12+\log(N/H)$ & 6.28$\pm$0.12 & 6.29$\pm$0.13 & 6.15$\pm$0.13 & 6.49$\pm$0.14 & 6.48$\pm$0.06 & 6.37$\pm$0.09\\
\hline
\end{tabular}
\end{center}
\end{minipage}
\end{table*}


We have calculated the ionic abundances of $O^+$, $O^{2+}$, $S^+$,
$S^{2+}$ and $N^+$ using standard expressions (Pagel et al. 1992), and
using for each ion its corresponding  line temperature, except for
[SII] in the objects for which no measurements of the [SII] $\lambda
\lambda$4068,4076 {\AA} lines exist, and [NII] for which no measurements of the
[NII]$\lambda$ 5755 {\AA} are reported. In theses cases, under the assumption of a
homogeneus electron temperature in the low excitation zone, we have
taken the aproximation of t([NII])$\approx$t([SII])$\approx$t([OII]). The computed ionic 
abundances relative to H$^+$ are also shown in Table 5.

\subsection{Ionisation correction factors and photo-ionisation models}

In order to derive the total chemical abundances we must add the
contribution of unseen stages of ionisation and this requires the
reconstruction of the ionisation structure of the region which, in turn,
allows the determination of appropriate ionisation correction factors
(ICF). To this purpose we have modelled each HII galaxy using
the recent version of the photo-ionisation code
CLOUDY (Ferland 2002). Each modelled galaxy has been
characterized by a set of input parameters including chemical
abundances, ionising continuum and nebular geometry in a very
simple way: the galaxy has been assumed to be spherically
symmetric with the ionised emmitting gas, assumed to be of constant
density, located at a distance very
large compared to its thickness, therefore resulting in a plane-parallel
geometry. Its degree of ionisation of the gas has been specified in terms of the 
ionisation parameter U, the ratio of the ionising photon density to the
particle density. This can be estimated from suitable line
ratios. At low metallicities, as the ones expected for HII galaxies,
the use of the sum of the [SII] lines at $\lambda\lambda$ 6717,6731 {\AA} can be used
({\em e.g.} Garc{\'\i}a-Vargas, Bressan \& D{\'\i}az 1995).

 The nebula has been assumed to be
ionised by a source represented by a single star whose effective temperature  has been
estimated from the $R_{23}$ and $R_{33}$ parameters as described in
Cervi{\~n}o \& Mas-Hesse (1994) and whose spectral energy distribution has
been taken from the  CoStar NLTE single-star stellar atmosphere models
(Schaerer and de Koter 1997) of 0.2 times the solar metallicity. Finally, as input
abundances we have used the sum of the respective ionic abundances for O and S,
the N/O ratio as given by the N$^+$/O$^+$ ratio and the rest of heavy
elements scaled to oxygen in solar proportions as given by Grevesse \&
Sauval (1998).
The refractory elements: Fe, Mg, Al, Ca, Na, Ni have been  depleted by a
factor of 10, and Si by a factor of 2 (Garnett et al. 1995), to take
into account the presence of dust grains. With these imput parametrers
we have calculated a first photo-ionisation model which has then been
optimized.

The predicted line intensities for the best fitting models, together
with the resulting functional parameters: U, T$_{eff}$ and chemical
abundances are given in Table 6 in comparison with observations. Also
listed are the ICF for S and N as well as their total abundance values.

\begin{table*}
\begin{minipage}{150mm}
\footnotesize
\caption{{\large Comparison between observations (Obs.) and results
    from photo-ionisation models, as described in the text.}}
\begin{center}
\begin{tabular}{c|cc|cc|cc|}
\hline
\hline
  & \multicolumn{2}{c}{II Zw 40} & \multicolumn{2}{c}{Mrk 5} & \multicolumn{2}{c}{0749+568} \\
 & Obs. & Model  & Obs. & Model  & Obs. & Model  \\
\hline

$\log U$ &  -2.23$\pm$0.08 & -2.23 & -2.6$\pm$0.2 & -2.85 & -2.55$\pm$0.10 & -2.71 \\
$T_{eff}(K)$   &          --   & 47300 &       --   & 50100 &           -- & 50400 \\

\hline
3727 [OII]  &  83.9$\pm$1.2  &  84.5 & 212.9$\pm$3.1 & 197.1 & 166.8$\pm$4.3 & 157.9\\
4363 [OIII] &  10.9$\pm$0.4  &  10.4 &   4.4$\pm$0.5 &   5.2 &   9.8$\pm$1.1 &   6.8\\
4959 [OIII] & 246.2$\pm$2.1  & 248.1 & 129.8$\pm$1.8 & 126.5 & 167.2$\pm$3.9 & 153.2\\
5007 [OIII] & 740.9$\pm$5.6  & 746.7 & 381.5$\pm$4.5 & 380.7 & 488.0$\pm$9.9 & 461.3\\
6312 [SIII] &   1.4$\pm$0.04  &   1.9 &   1.6$\pm$0.2 &   1.6 &   2.3$\pm$0.3 &   1.1\\
6720 [SII]  &  11.3$\pm$0.2  &  11.9 &  37.6$\pm$0.2 &  28.2 &  23.4$\pm$0.3 &  14.1\\
7325 [OII]  &   2.9$\pm$0.1  &   3.3 &   5.6$\pm$0.4 &   4.5 &   3.4$\pm$0.2 &   5.0\\
9069 [SIII] &  13.9$\pm$0.6  &  17.0 &  15.9$\pm$1.2 &  14.1 &  12.7$\pm$1.4 &   9.1\\
9532 [SIII] &  27.9$\pm$1.0  &  42.2 &  32.2$\pm$2.2 &  34.9 &       --    &  22.5\\
\hline
t([OII]) & 1.27$\pm$0.06 & 1.39 & 1.31$\pm$0.08 & 1.37 & 1.38$\pm$0.12 & 1.38 \\
t([OIII])& 1.34$\pm$0.03 & 1.31 & 1.22$\pm$0.06 & 1.30 & 1.54$\pm$0.10 & 1.34 \\
t([SIII])& 1.30$\pm$0.05 & 1.29 & 1.33$\pm$0.16 & 1.28 & 1.86$\pm$0.36 & 1.32 \\
\hline
12+log(O$^+$/H$^+$) & 7.08$\pm$0.07 & 7.15 & 7.45$\pm$0.10  & 7.45 & 
7.28$\pm$0.28 & 7.34 \\
12+log(O$^{2+}$/H$^+$) & 8.03$\pm$0.03 & 8.04 & 7.86$\pm$0.07 & 7.67 &
7.71$\pm$0.08 & 7.75 \\
12+log(O/H)        & 8.08$\pm$0.03 & 8.09 & 8.00$\pm$0.07 & 7.88 &
7.85$\pm$0.14 & 7.89  \\
\hline
12+log(N$^+$/H$^+$) & 5.77$\pm$0.04 & 5.85 & 6.09$\pm$0.05 & 6.26 & 5.75$\pm$
0.07 & 5.76 \\
ICF(N$^+$)          &-- & 9.13 &--& 3.04 &--& 3.90 \\
12+log(N/H)        & 6.73$\pm$0.04 & 6.81 & 6.58$\pm$0.05 & 6.75 & 6.34$\pm$0.07 & 6.35 \\
log(O$^+$/N$^+$)      & 1.31$\pm$0.04 & 1.29 & 1.29$\pm$0.05 & 1.19 & 1.48$\pm$0.08 & 1.59 \\   
\hline
12+log(S$^+$/H$^+$) & 5.18$\pm$0.05 & 5.38 & 5.68$\pm$0.19 & 5.77 & 5.43$\pm$
0.07 & 5.41 \\ 
12+log(S$^{2+}$/H$^+$) & 5.97$\pm$0.04 & 6.18 & 6.03$\pm$0.03 & 6.03 &
5.82$\pm$0.14 & 5.79 \\
12+log($S^++S^{2+}$/H$^+$) & 5.97$\pm$0.04 & 6.24 & 6.19$\pm$0.09 & 6.22 & 5.97$\pm$0.12 & 5.94 \\
ICF(S$^+$+S$^{2+}$)    &--& 1.24 &--& 0.86 &--& 0.92 \\
ICF(S$^+$)           &--& 9.09 &--& 2.41 &--& 3.09 \\
12+log(S/H)        & 6.13$\pm$0.04 & 6.36 & 6.12$\pm$0.13 & 6.24 & 5.93$\pm$0.12 & 5.97\\
\hline
\hline
\end{tabular}
\end{center}
\end{minipage}
\end{table*}


\begin{table*}
\begin{minipage}{150mm}
\footnotesize
\setcounter{table}{5}
\caption{{\large Continued}}
\begin{center}
\begin{tabular}{c|cc|cc|cc|}
\hline
\hline
  & \multicolumn{2}{c}{0926+606}& \multicolumn{2}{c}{Mrk 709} & \multicolumn{2}{c}{Mrk 22} \\
 & Obs. & Model & Obs. & Model & Obs. & Model \\
\hline
$\log U$ &  -2.6$\pm$0.1 & -2.72 & -2.6$\pm$0.1 & -2.81 & -2.45$\pm$0.10 & -2.59  \\
$T_{eff}$   &      --     & 50200 &      --    & 49400 &      --      & 48700 \\

\hline
3727 [OII]  & 178.5$\pm$4.2  & 171.3 & 183.6$\pm$18.1& 182.0 & 148.7$\pm$2.3 & 145.9 \\
4363 [OIII] &   8.3$\pm$0.3  &   6.4 &   8.8$\pm$0.5 &   5.4 &   8.2$\pm$0.3 &   7.5 \\
4959 [OIII] & 162.8$\pm$1.0  & 153.2 & 121.5$\pm$0.9 & 129.9 & 182.4$\pm$1.6 & 179.4 \\
5007 [OIII] & 477.2$\pm$2.6  & 461.1 & 369.6$\pm$4.2 & 391.1 & 545.5$\pm$4.2 & 539.9 \\
6312 [SIII] &   2.0$\pm$0.1  &   1.7 &   1.4$\pm$0.1 &   1.4 &   3.0$\pm$0.1 &   1.6 \\
6720 [SII]  &  32.0$\pm$0.5  &  24.3 &  62.3$\pm$0.4 &  23.7 &  26.6$\pm$0.5 &  18.9 \\
7325 [OII]  &   4.3$\pm$0.2  &   4.8 &   5.7$\pm$0.2 &   5.2 &   2.6$\pm$0.3 &   4.6 \\
9069 [SIII] &  12.2$\pm$1.7  &  15.0 &   8.7$\pm$1.0 &  12.7 &  14.8$\pm$2.0 &  14.4 \\
9532 [SIII] &  36.0$\pm$4.6  &  37.1 &       --    &  31.5 &      --     &  35.7 \\
\hline
t([OII]) & 1.23$\pm$0.04 & 1.37 & 1.50$\pm$0.16 & 1.39 & 1.16$\pm$0.09 & 1.36  \\
t([OIII])& 1.43$\pm$0.03 & 1.31 & 1.67$\pm$0.06 & 1.31 & 1.35$\pm$0.03 & 1.31  \\
t([SIII])& 1.52$\pm$0.18 & 1.29 & 1.65$\pm$0.23 & 1.29 & 1.98$\pm$0.28 & 1.29  \\
\hline
12+log(O$^+$/H$^+$) & 7.46$\pm$0.06 & 7.40 & 7.20$\pm$0.18 & 7.41 & 
7.49$\pm$0.13 & 7.34 \\
12+log(O$^{2+}$/H$^+$) & 7.77$\pm$0.03 & 7.77 & 7.49$\pm$0.04 & 7.68 &
7.89$\pm$0.03 & 7.86 \\
12+log(O/H)        & 7.94$\pm$0.04 & 7.92 & 7.68$\pm$0.05 & 7.87 &
8.04$\pm$0.06 & 7.98  \\
\hline
12+log(N$^+$/H$^+$) & 5.96$\pm$0.04 & 5.88 & 6.37$\pm$0.07 & 6.45 & 5.92$\pm$0.07 & 5.85 \\
ICF(N$^+$)          &-- & 3.80 &--& 3.26 &--& 4.56 \\
12+log(N/H)        & 6.54$\pm$0.04 & 6.46 & 6.88$\pm$0.07 & 6.96 &
6.58$\pm$0.07 & 6.51 \\
log(O$^+$/N$^+$)      & 1.43$\pm$0.04 & 1.52 & 0.81$\pm$0.03 & 0.97 & 1.47$\pm$0.07 & 1.49 \\   
\hline
12+log(S$^+$/H$^+$) & 5.83$\pm$0.15 & 5.71 & 5.79$\pm$0.03 & 5.64 & 5.85$\pm$0.19
 & 5.60 \\ 
12+log(S$^{2+}$/H$^+$) & 5.95$\pm$0.12 & 6.07 & 5.71$\pm$0.01 & 5.93 &
5.85$\pm$0.12 & 6.08 \\
12+log($S^++S^{2+}$/H$^+$) & 6.20$\pm$0.13 & 6.23 & 6.05$\pm$0.07 & 6.11 & 6.15$\pm$0.16 & 6.12 \\
ICF(S$^+$+S$^{2+}$)    &--& 0.92  &--& 0.87  &--& 0.99  \\
ICF(S$^+$)           &--& 3.04 &--& 2.58  &--& 3.92  \\
12+log(S/H)        & 6.16$\pm$0.13 & 6.26 & 5.99$\pm$0.02 & 6.14 & 6.15$\pm$0.16 & 6.25\\
\hline
\hline
\end{tabular}
\end{center}
\end{minipage}
\end{table*}

\begin{table*}
\begin{minipage}{150mm}
\footnotesize
\setcounter{table}{5}
\caption{{\large Continued}}
\begin{center}
\begin{tabular}{c|cc|cc|cc|}
\hline
\hline
  & \multicolumn{2}{c}{Mrk 1434}&  \multicolumn{2}{c}{Mrk 36} & \multicolumn{2}{c}{VII Zw 403} \\ 
 & Obs. & Model & Obs. & Model & Obs. & Model \\
\hline
$\log U$ & -2.4$\pm$0.1 & -2.36 & -2.65$\pm$0.20 & -2.51 & -2.6 $\pm$0.1 & -2.62 \\
$T_{eff}$   &      --    & 46000 &      --      & 42300 &        --   & 42500 \\

\hline
3727 [OII]  &  96.8$\pm$0.6  &  97.3 & 129.3$\pm$1.5 & 124.1 & 133.3$\pm$0.9  & 132.3\\
4363 [OIII] &  10.4$\pm$0.2  &   7.7 &   9.6$\pm$0.5 & 6.9   &   7.1$\pm$0.2  &   4.4\\
4959 [OIII] & 170.4$\pm$0.8  & 171.7 & 162.2$\pm$1.6 & 170.7 & 114.3$\pm$0.8  & 117.0\\
5007 [OIII] & 502.8$\pm$2.1  & 516.9 & 483.4$\pm$4.2 & 513.7 & 345.5$\pm$3.9  & 352.0\\
6312 [SIII] &   1.6$\pm$0.1  &   0.8 &   1.8$\pm$0.2 & 1.8   &   1.3$\pm$0.1  &   1.3\\
6720 [SII]  &  15.1$\pm$0.2  &   6.8 &  20.0$\pm$0.2 & 19.6  &  13.0$\pm$0.2  &  14.4\\
7325 [OII]  &   2.3$\pm$0.2  &   2.7 &   3.9$\pm$0.4 & 3.8   &   2.2$\pm$0.2  &   3.5\\
9069 [SIII] &   9.2$\pm$0.8  &   6.9 &  11.9$\pm$0.4 & 16.6  &  11.7$\pm$1.3  &  11.9\\
9532 [SIII] &      --      &  17.1 &  29.5$\pm$3.5 & 41.2  &     --       &  29.4\\
\hline
t([OII]) & 1.24$\pm$0.08 & 1.38 & 1.37$\pm$0.12 & 1.37 & 1.41$\pm$0.12 & 1.33 \\
t([OIII])& 1.55$\pm$0.02 & 1.34 & 1.53$\pm$0.05 & 1.29 & 1.52$\pm$0.03 & 1.25 \\
t([SIII])& 1.73$\pm$0.20 & 1.32 & 1.62$\pm$0.30 & 1.27 & 1.30$\pm$0.14 & 1.24 \\
\hline
12+log(O$^+$/H$^+$) &  7.20$\pm$0.10 & 7.00 & 7.18$\pm$0.12  & 7.28 & 
7.15$\pm$0.11 & 7.32 \\
12+log(O$^{2+}$/H$^+$) & 7.71$\pm$0.02 & 7.70 & 7.70$\pm$0.04 & 7.87 &
7.56$\pm$0.02 & 7.73 \\
12+log(O/H)        & 7.83$\pm$0.04 & 7.78 & 7.81$\pm$0.06 & 7.97 &
7.70$\pm$0.05 & 7.88  \\
\hline
12+log(N$^+$/H$^+$) & 5.58$\pm$0.08 & 5.38 & 5.58$\pm$0.14 & 5.59 & 5.56$\pm$0.09 & 5.58 \\
ICF(N$^+$)          &-- & 6.96 &--& 5.23 &--& 4.24 \\
12+log(N/H)        & 6.42$\pm$0.08 & 6.22 & 6.30$\pm$0.14 & 6.31 & 6.18$\pm$0.09 & 6.21 \\
log(O$^+$/N$^+$)      & 1.58$\pm$0.08 & 1.62 & 1.52$\pm$0.07 & 1.68 & 1.56$\pm$0.08 & 1.74 \\   
\hline
12+log(S$^+$/H$^+$) & 5.18$\pm$0.17 & 5.15 & 5.63$\pm$0.27 & 5.63 & 5.48$\pm$0.29
 & 5.51 \\ 
12+log(S$^{2+}$/H$^+$) & 5.71$\pm$0.09 & 5.76 & 5.86$\pm$0.15 & 6.16 &
5.96$\pm$0.12 & 6.02 \\
12+log($S^++S^{2+}$/H$^+$) & 5.82$\pm$0.11 & 5.85 & 6.06$\pm$0.20 & 6.00 & 6.09$\pm$0.16 & 6.13 \\
ICF(S$^+$+S$^{2+}$)    &--& 1.07 &--& 1.01 &--& 0.95 \\
ICF(S$^+$)           &--& 5.43 &--& 4.37 &--& 4.02 \\
12+log(S/H)        & 5.85$\pm$0.11 & 5.92 & 6.07$\pm$0.20 & 6.32 & 6.06$\pm$0.16 & 6.16\\
\hline
\hline
\end{tabular}
\end{center}
\end{minipage}
\end{table*}


\begin{table*}
\begin{minipage}{150mm}
\footnotesize
\setcounter{table}{5}
\caption{{\large Continued}}
\begin{center}
\begin{tabular}{c|cc|cc|cc|}
\hline
\hline
  & \multicolumn{2}{c}{UM 461} & \multicolumn{2}{c}{UM 462}&  \multicolumn{2}{c}{Mrk 209} \\
 & Obs. & Model & Obs. & Model & Obs. & Model \\
\hline
$\log U$  & -2.1$\pm$0.1 & -2.15 & -2.6$\pm$0.1 & -2.65 & -2.3$\pm$0.1 & -2.29 \\
$T_{eff}$    &      --    & 46600 &     --     & 48000 &     --     & 43200 \\

\hline
3727 [OII]  &  52.7$\pm$1.5 &  57.7 & 174.2$\pm$1.0  & 170.0 &  71.9$\pm$0.2 &  68.7 \\
4363 [OIII] &  13.6$\pm$1.7 &  11.3 &   7.8$\pm$0.2  & 7.5   &  12.7$\pm$0.1 &   9.6 \\
4959 [OIII] & 203.9$\pm$3.4 & 208.8 & 166.3$\pm$0.9  & 173.1 & 196.0$\pm$0.3 & 200.8 \\
5007 [OIII] & 602.2$\pm$9.0 & 628.4 & 492.9$\pm$2.3  & 521.0 & 554.3$\pm$0.8 & 604.4 \\
6312 [SIII] &   2.0$\pm$0.1 &   0.8 &   2.1$\pm$0.2  & 2.0   &   1.7$\pm$0.1 &   1.1 \\
6720 [SII]  &  10.3$\pm$0.2 &   4.4 &  28.3$\pm$0.5  & 24.0  &  10.6$\pm$0.4 &   6.8 \\
7325 [OII]  &   2.1$\pm$0.2 &   1.9 &   4.3$\pm$0.2  & 5.3   &   1.7$\pm$0.2 &   2.4 \\
9069 [SIII] &  12.4$\pm$0.8 &   6.2 &  10.5$\pm$1.5  & 17.1  &  12.2$\pm$1.2 &   8.7 \\
9532 [SIII] &  21.7$\pm$0.7 &  15.3 &  35.0$\pm$3.6  & 42.5  &       --    &  21.7 \\
\hline
t([OII]) & 1.66$\pm$0.19 & 1.49 & 1.19$\pm$0.03 & 1.38 & 1.23$\pm$0.08 & 1.40 \\
t([OIII])& 1.62$\pm$0.05 & 1.46 & 1.38$\pm$0.02 & 1.33 & 1.62$\pm$0.01 & 1.38 \\
t([SIII])& 1.95$\pm$0.16 & 1.43 & 1.66$\pm$0.25 & 1.31 & 1.60$\pm$0.17 & 1.36 \\
\hline
12+log(O$^+$/H$^+$) & 6.60$\pm$0.10 & 6.85 & 7.51$\pm$0.04 & 7.36 & 
7.08$\pm$0.10 & 7.00 \\
12+log(O$^{2+}$/H$^+$) & 7.74$\pm$0.04 & 7.85 & 7.82$\pm$0.02 & 7.83 &
7.71$\pm$0.01 & 7.89 \\
12+log(O/H)        & 7.77$\pm$0.04 & 7.89 & 7.99$\pm$0.03 & 7.95 &
7.80$\pm$0.03 & 7.94  \\
\hline
12+log(N$^+$/H$^+$) & 5.26$\pm$0.11 & 5.46 & 5.91$\pm$0.04 & 5.84 & 5.58$\pm$0.07 & 5.53 \\
ICF(N$^+$)          &-- & 10.86 &--& 4.09 &--& 8.79 \\
12+log(N/H)        & 6.29$\pm$0.11 & 6.50 & 6.53$\pm$0.04 & 6.45 & 6.52$\pm$0.07 & 6.48 \\
log(O$^+$/N$^+$)      & 1.32$\pm$0.09 & 1.39 & 1.51$\pm$0.09 & 1.52 & 1.42$\pm$0.06 & 1.47 \\   
\hline
12+log(S$^+$/H$^+$) & 4.95$\pm$0.09 & 4.90 & 5.83$\pm$0.17 & 5.65 & 5.18$\pm$0.10
 & 5.15 \\ 
12+log(S$^{2+}$/H$^+$) & 5.67$\pm$0.05 & 5.68 & 5.88$\pm$0.12 & 6.14 &
5.86$\pm$0.10 & 5.86 \\
12+log($S^++S^{2+}$/H$^+$) & 5.75$\pm$0.06 & 5.75  & 6.16$\pm$0.15 & 6.26 & 5.95$\pm$0.10 & 5.93 \\
ICF(S$^+$+S$^{2+}$)    &--& 1.35 &--& 0.98 &--& 1.22 \\
ICF(S$^+$)           &--& 9.48 &--& 4.00 &--& 7.52 \\
12+log(S/H)        & 5.88$\pm$0.06 & 5.90 & 6.15$\pm$0.15 & 6.30 & 6.03$\pm$0.10 & 6.05\\
\hline
\hline
\end{tabular}
\end{center}
\end{minipage}
\end{table*}

\section{Discussion}

\subsection{Line temperatures}
To compute chemical abundances in ionised gas nebulae, knowledge of  
the electron temperature is required. Since the assumption of an
isothermal gas region may not be valid due to temperature 
fluctuations (Peimbert 1967) and differences between 
the various line temperatures in each internal zone (Garnett 1992), we must 
use the appropriate line temperature for the calculation of the
abundance of each ion. These temperatures can be deduced directly from the
corresponding ratio of emission lines, but usually not all these lines are 
accesible in the spectra or have large associated errors. In these 
cases,  some assumptions are usually adopted about the temperature
structure through the nebula. For HII galaxies it is custommary to
assume a two-zone model  with a low ionisation zone where the [OII],
[NII], [NeII] and [SII] lines are formed, and a high ionisation zone in which
the [OIII], [NeIII] and  [SIII] lines are formed;
photo-ionisation models are then used to relate the temperatures representative of each zone (see, for example Pagel et al. 1992). In some cases, an
intermediate zone is assumed as in Garnett (1992) for the S$^{++}$ ion. 

The fact that we have been able to measure line temperatures which, in
principle, correspond to these three ionisation zones, allows us to
investigate to which extent these assumptions hold for the case of HII
galaxies. Figure 3 shows a comparison between t([OII]) and t([SII]) for our
observed objects and published data on HII galaxies for which we have
derived the line temperatures in the way described in section 3.2. 
The 1:1 relation is shown as a dashed-dotted line. Two sequences of
models of the same kind
as those described in section 3.3, but with different
values of the density: 10, 100 cm$^{-3}$ are also shown. 
Most of the data
are below the 1:1 relation and are better represented, at least for our
observed objects, by the sequence of
models with n$_e$ = 100 cm$^{-3}$ thus reflecting the density
dependence of the two involved temperatures.
However, for the whole HII galaxy sample, the average t([SII]) is still
lower than predicted by models. At any rate, since for this kind of
objects S$^{2+}$ is the
main contributor to the total S abundance, the commonly taken assumption 
of t([OII])=t([SII]) would not greatly affect the total abundance
determination.

\begin{figure}
\setcounter{figure}{2}
\begin{minipage}{85mm}
\psfig{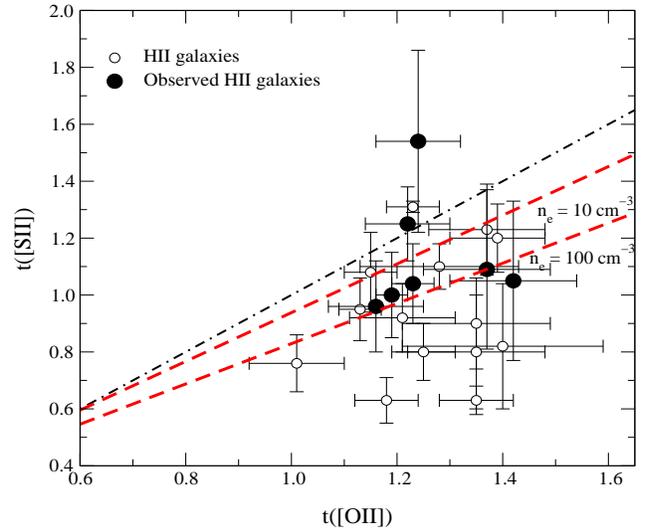}
\caption{A comparison between the measured line temperatures of
$[SII]$ and $[OII]$. Solid symbols correspond to the data in this
paper. Open symbols correspond to data by Izotov et al.(1994), Izotov et al.(1997) 
and Izotov \& Thuan (1998). The 1:1 relation is shown as a dashed-dotted
line. Two sequences of models computed with electron densities of 10
and 100 cm$^{-3}$, as labelled, are also shown. }
\end{minipage}
\end{figure} 

  The relation between t([OII]) and t([OIII]) is
shown in Figure 4. We have also added to this plot  values of t([OII]) and
t([OIII]) 
calculated from published data on HII galaxies (Izotov et al. 1994,
1997; Izotov \& Thuan 1998; Guseva et al. 2000).
Also shown are two relations commonly
used for abundance determinations, based on photo-ionisation models from
Stasi{\'n}ska (1980)  (solid line)
\[ t([OII]) = 0.7 t([OIII])+0.3\]
and Stasi{\'n}ska (1990) (broken line)
\[ t([OII])^{-1}=0.5 (t([OIII])^{-1}+0.8)\] 

\begin{figure}
\setcounter{figure}{3}
\begin{minipage}{85mm}
\psfig{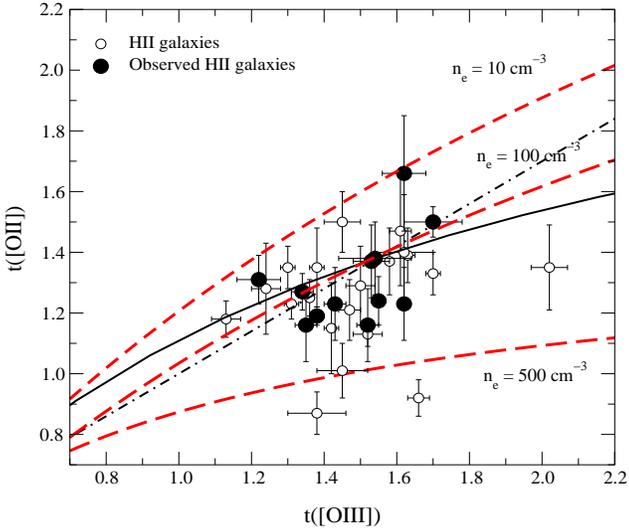}
\caption{Comparison between the measured line temperatures of
$[OIII]$ and $[OII]$. Solid symbols correspond to the data in this
paper. Open symbols correspond to similar data of HII galaxies by
Izotov et al.(1994), Izotov et al.(1997) 
and Izotov \& Thuan (1998). The lines are the model-deduced relations
for these line temperatures. The solid line corrresponds to the models by
Stasi{\'n}ska (1980). The broken one corresponds to those by 
Stasi{\'n}ska (1990). }
\end{minipage}
\end{figure} 

Our HII galaxies, and most of those in the literature, show t([OIII]) values
between 1.2 and 1.8 and, within the errors, cluster around the model fits. The
data point with the highest t([OIII]), about 2.1, shows a better agreement with the
second sequence of models. There are some data points, however, that
lie significantly below the theoretical relation. In this metallicity
regime, large scale temperature fluctuations that would yield t([OIII])
values higher than the corresponding ion weighted mean temperature are
not predicted (Garnett 1992), therefore the position of these points
correspond to objects that show values of t([OII]) which are lower than
predicted by the theoretical sequences. They
all belong to the samples by Izotov and co-workers and correspond
to: SBS0940+544N, SBS0832+699, SBS1533+574A and SBS0741+535. This latter
object has a density  of the order of 500
cm$^{-3}$, higher than the typical value of about 100 cm$^{-3}$. Given
the density dependence of t([OII]), this could explain the position of this
point in the diagram. Three sequences of models computed with different
values of the density: 10, 100 and 500 cm$^{-3}$ are also shown
in Fig 4 . Our model sequence for n$_e$ = 100 cm$^{-3}$ agrees very
well with the fit to the models of Stasi\'nska (1990) and the sequence at
n$_e$ = 500 cm$^{-3}$ reproduces well the data for SBS0741+535. This
explanation, however, is, in principle,  not valid for the
rest of the objects with low t([OII]) values for which the derived
electron densities are close to the typical one.  Uncertainties in
reddening could also be invoked. However, in order to reconcile
derived and model predicted values of t([OII]) the reddening constant
should have been overestimated by an amount unplausibly large (about 1
dex). Therefore these objects require further investigation. 

The adopted value of t([OII]) affects the computed ionic abundance of
O$^+$ and can become important in objects of relatively low excitation.
For SBS0832+699, SBS1533+574A and SBS0741+535, the use of the derived
t([OII]) instead of the value predicted by the Stasi\'nska (1990) relation
translates into higher O$^+$/H$^+$ abundances by factors of
5.2, 2.8 and 4.7 respectively and total oxygen
abundances higher by 0.33, 0.21 and 0.38 dex, values much larger than the
errors quoted by Izotov and co-workers (0.02, 0.03 and
0.02). Therefore, these kind of effects should be taken into account if
accurate values of abundances for these objects are sought. In the
absence of directly derived [OII] temperatures, higher excitation
objects, where the O$^+$/H$^+$ fraction contributes less to the total
oxygen abundance,  should provide more accurate abundance determinations.

Regarding [SIII], Garnett (1992), from data on HII galaxies and GEHR,
suggested that t([SIII]) is in fact intermediate between t([OII]) and
t([OIII]) and provided an empirical fit:
\[ t(S^{2+})=0.83 t(O^{2+})+0.17\] 
where t(S$^{2+}$) and t(S$^{2+}$) denote ion-weighted mean temperatures.
The relation between both temperatures, t([OIII]) and t([SIII]) is shown in Figure
5 both for our observed galaxies and those of Garnett (1989), together with
the empirical fit above. It can be seen in the figure that,  although this
fit works rather well when compared
with the temperatures deduced directly from observations, there are
non-negligible deviations that might affect the calculation  of sulphur
abundances in the regime of high excitation. Therefore, the validity of
this relation should be further explored. It should also be taken into
account that the atomic coefficients for [SIII] are suffering
continuous revisions which may significantly affect the derived values
of t([SIII]). This is illustrated in Table 7 where these temperatues are
listed for the observed objects using different sets of coefficients.


\begin{table}
\begin{minipage}{83mm}
\footnotesize
\setcounter{table}{6}
\caption{t([SIII]) from different sets of atomic coefficients}
\begin{center}
\begin{tabular}{lccc}
\hline
\hline
Refs \footnote{References: GMZ95:{\em Galavis, Mendoza \& Zeippen 1995}, T97:{\em Tayal 1997}, TG99:{\em Tayal \& Gupta 1999.}} & GMZ95 & T97 & TG99 \\
\hline
II Zw 40   & 1.47 & 1.21 & 1.30 \\ 
Mrk 5    & 1.51 & 1.26 & 1.33 \\
0749+568 & 2.16 & 1.74 & 1.86 \\
0926+606 & 1.74 & 1.42 & 1.52 \\
Mrk 709  & 1.91 & 1.57 & 1.65 \\
Mrk 22   & 2.32 & 1.85 & 1.98 \\
Mrk 1434 & 1.97 & 1.61 & 1.73 \\
Mrk 36   & 1.85 & 1.53 & 1.62 \\
VII Zw 403 & 1.47 & 1.25 & 1.30 \\
UM 461   & 1.98 & 1.84 & 1.95 \\
UM 462   & 1.92 & 1.53 & 1.66 \\
Mrk 209  & 1.84 & 1.52 & 1.60 \\
\hline
\end{tabular}
\end{center}
\end{minipage}
\end{table}

\begin{figure}
\setcounter{figure}{4}
\begin{minipage}{85mm}
\psfig{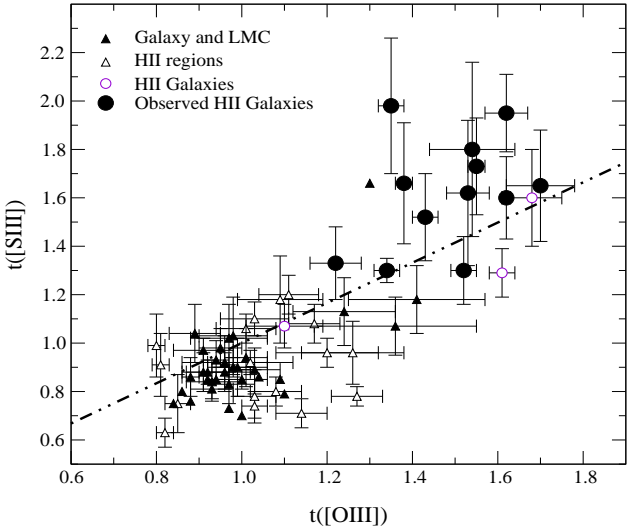}
\caption{A comparison between the measured line temperatures of
$[OIII]$ and $[SIII]$. The dark dot-dashed line is the model-deduced 
relation for the same ion-weighted temperatures from Garnett 1992 }
\end{minipage}
\end{figure} 

\subsection{Ionic fractions}


The fact that we can derive the ionic fractions of O, N and S allows to probe
the ionisation structure of the observed nebulae and test the validity of some
commonly adopted assumptions. The first of them relates to the equality of the
fractions of neutral hydrogen and oxygen:
\[  \frac{O^0}{O} = \frac{H^0}{H} \]
that allows to calculate O/H as:
\[  \frac{O}{H} = \frac{O^+ + O^{2+}}{H^+} \]
For our objects, the best fitting models yield values of O$^0$/O and H$^0$/H
which are equal inside 95 \%  except for two objects: IIZw40 and Mrk209 in
which the ratio 
deviate from unity by 20 \%  and 10 \%  respectively. These are the objects with
the highest densities in our sample: 290 cm$^{-3}$ for IIZw40 and 190 cm$^{-3}$ for
Mrk209.   In any case, given the low proportion of neutral oxygen in
these objects, the assumption above seems to be well justified. 

Also the aproximation $(N/O) \approx (N^+/O^+)$, seems to hold to better than
85 \%  and the N/O ratio, as computed using the ICFs for
nitrogen from the models, equals the N$^+$/O$^+$ derived
observationally except for one object : Mrk 5 for which this latter
ratio is found to be larger by about 25 \%. 


The situation regarding sulphur requires further attention. Even when
we observe the two major ionisation states of sulphur: S$^+$ and
S$^{++}$, we still have to correct for the possible presence of
S$^{3+}$ which, in the absence of IR observations of [SIV], means we
have to rely on photo-ionisation models. 

A first scheme for the ionisation correction necessary to calculate the
total abundance of sulphur was given by Peimbert \& Costero (1969) based
on the similarity of the ionisation potentials of S$^{2+}$ and
O$^{+}$. Then: S/O = (S$^+$ + S$^{++}$)/O$^+$
and 
 \[ ICF(S^++S^{2+})=\frac{S/H}{(S^++S^{2+})/H^+}=
 \left(\frac{S}{S^++S^{2+}}\right)\left(\frac{H^+}{H}\right) = \]
 \[ =\left(\frac{O}{O^+}\right)\left(\frac{H^+}{H}\right) \] 

This relation, however, overestimates S/H for regions of high
excitation ({\em e.g.} Barker 1978; Pagel 1980). Modified versions
of it, based on photo-ionisation models by Stasi\'nska
(1978), have been proposed by Barker (1980) and French (1981). 
All of them can be written as:
 \[ ICF(S^++S^{2+})= \left[ 1 - \left(1 -
     \frac{O^+}{O}\right)^{\alpha}\right]^{-1/\alpha} \]
where $\alpha$ takes the value 3, 2 or 1 in the approximations by Barker
(1980), French (1981) and Peimbert \& Costero (1969) respectively.  
More recently, Izotov et al. (1994), based on
photo-ionisation models by Stasi\'nska (1990), use a polynomial fitting
with O$^+$/O as the independent variable, which is very close to the
expresion above for  $\alpha$ = 2. 

Figure 6 shows the ICFs for sulphur: ICF(S$^+$ + S$^{2+}$) as a
function of O$^+$/O for different values of $\alpha$ together with the
values derived by our individual models for the observed objects. For
most objects, the obtained ICF are close to the above experession for
$\alpha$ = 3. However, the comparison with the only observational datum,
comming from IR observations of the [SIV]$\lambda$ 10.5 $\mu$m by Nollenberg et
al. (2002) lies nicely on the curve of $\alpha$ = 2. Obviously, more data
are needed in order to assess the validity of our models.


\begin{figure}
\setcounter{figure}{5}
\begin{minipage}{85mm}
\psfig{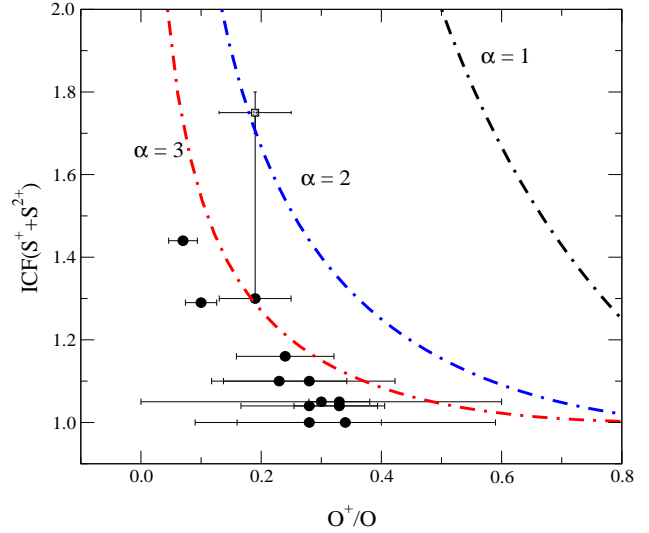}
\caption{Ionisation correction factors, derived from our tailored
  photo-ionisation models for the objects in our sample. The continuous 
line represents the predictions by the scheme of Peimbert \& Costero
(1978). The dashed-dotted lines corresponds to the fits to Stasi\'nska
(1978) models of Barker (1980) for $\alpha$=3 and $\alpha$=2, as
labelled. Finally, the dashed line represents the fit to Stasi\'nska
(1990) models by Izotov et al. (1994).}
\end{minipage}
\end{figure} 

Another approach has been suggested by Mathis \& Rosa (1991) who propose the
use of both O$^+$/O$^{++}$ and S$^+$/S$^{++}$ to derive the ICFs for
various elements exploiting the fact that models which provide given
values of these two ratios have similar ICFs, even if they have
different geometry, chemical composition or ionising source. They then
derive the various ICFs as a polynomial fitting to these ratios.


\begin{figure}
\setcounter{figure}{6}
\begin{minipage}{85mm}
\psfig{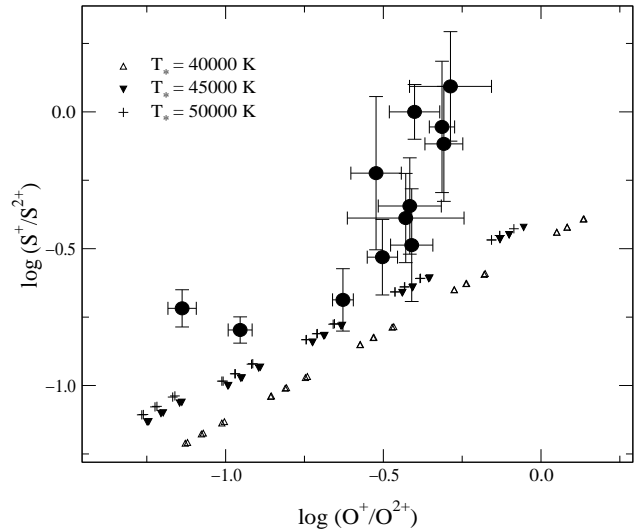}
\caption{The S$^+$/S$^{2+}$ vs O$^+$/O$^{2+}$ diagram for the observed
  HII galaxies as compared with results from photo-ionisation models
  (see text for details).}
\end{minipage}
\end{figure} 

When trying to use this method, the first question to ask is to
what extent the model sequences reproduce the ionisation structure
of the observed HII galaxies.  We can check this by comparing the
observationally derived ionic fractions and those computed by
models. Figure 7 shows the S$^+$/S$^{2+}$ vs O$^+$/O$^{2+}$ values as derived from
observations as compared to those predicted by our models for the objects
of the sample. Models with Z/Z$_{\odot}$ 0.05 and 0.1 and 0.2 are
shown. Open triangles, filled triangles and crosses correspond to
stellar effective temperatures of 40,000, 45,000 and 50,000 K and the
ionisation parameter varies between 10 $^{-3}$ and 10$^{-2}$ in steps
of 0.25 dex. As we can see, models distribute along a narrow diagonal
band. This is to be expected since the quotient of
these two ionic fractions is the ``eta parameter'' of V\'\i lchez \&
Pagel(1988) and constitutes a measure of the ``softness'' of the
ionising radiation. Therefore, nebula ionised by sources with similar
spectral energy distribution should lie along a straight line in the
diagram. The observed values are adequatelly reproduced by the
models, given the large errors involved, for the objects of intermediate
excitation. Significant deviations are however found for both high
excitation objects, to the left of the diagram, and low excitation
ones, to the right, in both cases in the sense of the models predicting
lower values of S$^+$/S$^{2+}$ than observed. Figure 8 shows a
comparison of the total sulphur abundances computed with the ICFs of
Mathis \& Rosa (1991) and those found in this work. In general, Mathis \&
Rosa's provides S/H values systematically larger than ours by about
0.15 dex which is inside the quoted uncertainty. Two objects however
deviate significantly: IIZw40 and UM461 which are the objects with the
highest excitation in our sample.


\begin{figure}
\setcounter{figure}{7}
\begin{minipage}{85mm}
\psfig{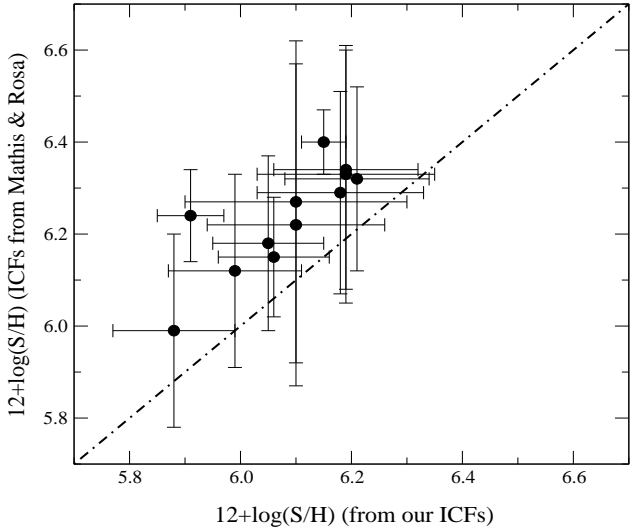}
\caption{A comparison of the total sulphur abundances derived using
  Mathis \& Rosa (1991) and our Ionisation correction factors. The
  dashed-dotted line represents the 1:1 relation.}
\end{minipage}
\end{figure} 

\subsection{Total abundances}


The relation between the metallicity, represented by 12+log(O/H), and 
the N/O  abundance ratio, assuming N/O = N$^+$/O$^+$ is plotted in
Figure 9 for the observed sample of objets (solid circles) together
with data corresponding to HII galaxies in the literature (open
circles). Two locations for the sun are shown in the diagram, the one
with the lowest N/O ratio corresponds to the data by Anders \& Grevesse
(1989). The other solar symbol corresponds to the more
recent data of Allende Prieto et al. (2001) for oxygen  and Holweger (2001)  for 
nitrogen. Most observed objects cluster around a constant value of
log(N/O) = -1.5. A couple of objects however show N/O ratios larger
than expected for their low oxygen abundances. The most deviating datum
corresponds to Mrk709 which shows an N/O ratio close to solar and an
oxygen abundance of about one tenth solar. Our emission line data agree well with
those of Terlevich et al. (1991) and the derived t$_e$([OIII]) and
t$_e$([SIII]) are equal within the errors, which suggests that the
oxygen abundance is probably well derived. This particular object, however, also shows
relatively intense lines of [OI] $\lambda$ 6300 \AA{} and [SII] $\lambda\lambda$ 6716, 6731 \AA{}
and has the lowest value of the equivalent width of H$\beta$ in our sample. Therefore it
might constitute a case for a relatively evolved galaxy in which the
contribution by shocks originated  from stellar winds and/or supernovae
is important. The large N/O abundance derived could be due to the
enhancement of the [NII]$\lambda\lambda$ 6548, 6584 \AA{} by the shock contribution.


\begin{figure}
\setcounter{figure}{8}
\begin{minipage}{85mm}
\psfig{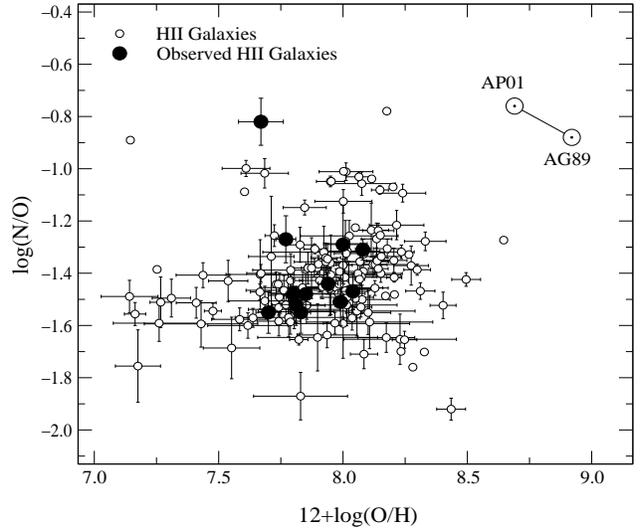}
\caption{The N/O ratio as a function of total oxyen abundance for the
  objects in our sample and HII galaxy data in the literature.}
\end{minipage}
\end{figure} 


 The relation between the metallicity, represented by 12+log(O/H), and 
the S/O  abundance ratio is plotted in Figure 10
for our sample objects and  HII galaxies compiled from the
literature. Solid circles represent the data of this work while solid squares
circles correspond to data by Skillman \& Kennicutt (1993) on IZw18 and
Skillman et al. (1994) on UGC4483. In both cases the
S$^{2+}$ abundances have been derived from the near infrared [SIII]
lines. We have added to the graph data by Izotov et al. (1994, 1997)
and Izotov \& Thuan (1998)
represented by open circles, where the S$^{2+}$ abundances have been derived
from the weak [SIII]$\lambda$ 6312 \AA{} line. In this latter case we have
recomputed the abundances to take properly into account the
observational errors and we have applied the sulphur ICFs calculated
from Barker (1980) which fits best our observations. 

Again two solar symbols are shown in
the diagram, one corresponding to the data by Anders \& Grevesse (1989)
and a higher one in which the more recent value of the solar oxygen
abundance by Allende Prieto et al. (2001) is adopted. In this case, all
the observed objects show S/O ratios which are below the new S/O solar
ratio, suggesting that some revision of the solar sulphur abundance
might be needed. 
The sources of observational errors in the measurement of the auroral lines and 
the uncertainties due to the modelling  of the unobserved properties 
of the nebulae lead to recognized difficulties in ascertaining any real trends of 
the relation between S/O and O/H. The data are inconclusive and keep some open
questions about the behaviour of the S/O ratio in the regime of low
metallicity. More good quality data  are obviously needed reaching the
near IR in order to explore this behaviour which can convey important
information about the nucleosynthesis processes in high mass stars and the initial mass
function at the high mass end. Data in the mid-IR to obtain [SIV]
intensities would also help to clarify this question.


\begin{figure}
\setcounter{figure}{9}
\begin{minipage}{85mm}
\psfig{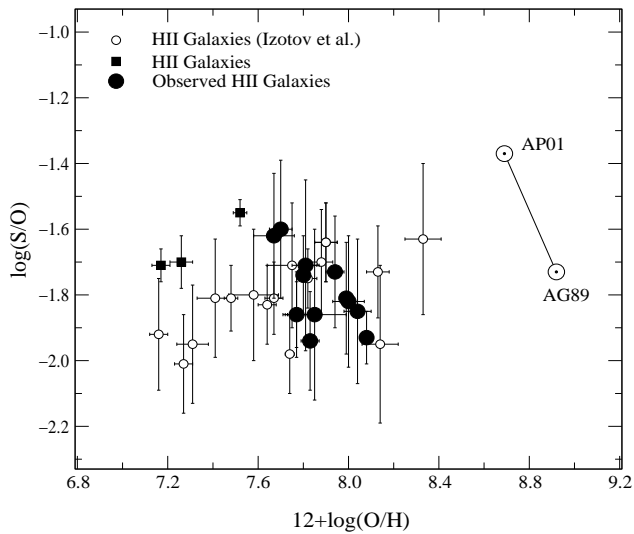}
\caption{The S/O ratio as a function of total oxyen abundance for the
  objects in our sample and HII galaxy data in the literature.}
\end{minipage}
\end{figure} 


Finally, we have added to the empirical calibration of the metallicity 
parameter, S$_{23}$= ([SII]+[SIII])/H$\beta$,   (D{\'\i}az \& P{\'e}rez-Montero, 2000) our
observed HII galaxies and also recent data from Kennicutt \& Garnett (2000) 
corresponding to HII regions in the Galaxy and the Magellanic Clouds; 
Oey et al. (2000) corresponding to  HII regions in the LMC; and the
sample of HII
regions 
by  Castellanos et al. (2002). This calibration, although showing a
substantial scatter, has the potentiallity of remaining single-valued
up to solar metallicity. This is specially important in the case of HII
galaxies most of which show values of the commonly used oxygen
abundance parameter R$_{23}$ in the turn-over
region of the abundance calibration. In fact, this precludes the
knowledge of the true abundance distribution of HII galaxies. However,
more data points are needed in order to reduce the scatter of the
relation and fill the existing gap at the low abundance end.

\begin{figure}
\setcounter{figure}{9}
\begin{minipage}{85mm}
\psfig{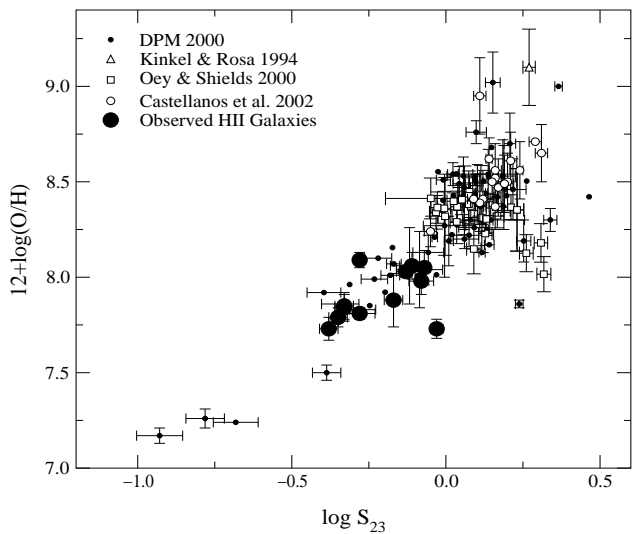}
\caption{The parameter $S_{23}$ plotted against the metallicity of the regions}
\end{minipage}
\end{figure} 

\section{Conclusions}
We have analyzed long-slit spectrophotometric observations of a sample
of 12 HII galaxies in the red and far red including the auroral and
nebular [SIII] lines at $\lambda$ 6312 \AA{} and $\lambda\lambda$ 9069 and 9532 \AA{}
respectively and the [OII] lines at $\lambda\lambda$ 7319 and 7330 \AA. For all the
objects, previous data in the blue-visible
exist and it has therefore been possible to derive directly at least
three line temperatures: t([OII]), t([OIII]) and t([SIII]). For 7 of the 12
observed objects it has also been possible to measure  t([SII]). The
temperatures for [OII] and [SII] are found to be representative of the
same zone. Regarding t([OII]) and t([OIII]), their values cluster around
those predicted by simple photo-ionisation models with intermediate
density (100 cm$^{-3}$). However, some objects show values of t([OII])
which are significantly below the theoretical relation. This could in
principle be due to a higher value of the density that affects the
derivation of t([OII]). The fact of using a t([OII]), representative of the
low ionisation zone, computed from photo-ionisation model sequences
instead of that directly derived, can
lead to underestimate the O$^+$/H$^+$ ionic abundance and hence the
total oxygen abundance by factors much larger than quoted observational
errors.
Regarding t([SIII]) most of the observed objects show values which are
slightly larger than those predicted from t([OIII]) by the linear fit given by
Garnett (1989) on the base of photo-ionisation models. The effect can be
significant for the high excitation objects.

For all the objects in the sample we have derived the ionic fractions
of O, N and S and have computed ICF for nitrogen and sulphur from
detailed photo-ionisation model fits to the data. We have confirmed the 
validity of the commonly taken
assumptions of O/H = (O$^+$+O$^{++}$)/H$^+$ and N/O =
N$^+$/O$^+$. Regarding sulphur, our ICFs for (S$^{+}$+S$^{++}$) are best reproduced by
Barker (1980) approximation formula. On the other hand, the ionisation
structure of the observed regions are not adequatelly reproduced by
theoretical models for relatively low exctitation objects and the total
sulphur abundances as computed with the ICF calculated from the method
of Mathis \& Rosa (1991) based on this ionisation structure are
systematically larger than those found in this paper.

Finally, regarding total abundances, the N/O ratio for our observed
objects, and also for a large sample of HII galaxies in the literature,
shows values below solar, although it is far from being constant. In
fact it spans almost an order of magnitude, with some objects having an
N/O ratio too high for their derived oxygen abundance. For the S/O
 observational errors are still too high to ascertain the
reality of any existing trend. 

\section*{Acknowledgements}
We would like to thank M. Castellanos, E. Terlevich, J.M. V\'\i lchez, C. Esteban, E. P\'erez and
D. Valls-Gabaud for very interesting dicussions and suggestions and
J. Zamorano for his help during the first observation campaign.\\
The INT is operated in the island of La Palma by the Isaac Newton Group
in the Spanish Observatorio del Roque de los Muchachos of the Instituto
de Astrof\'\i sica de Canarias. We thank CAT for awarding observing time.\\
This work has been partially supported by DGICYT project AYA-2000-0973.

\end{document}